\documentclass[showpacs,superscriptaddress,preprintnumbers]{revtex4}
\usepackage{array}
\usepackage{booktabs}
\usepackage{amsfonts}
\usepackage{amssymb}
\usepackage{amsmath}
\usepackage{epsfig}
\usepackage{array}
\usepackage{color}
\usepackage{subfig}
\usepackage{float}
\allowdisplaybreaks

\begin{document}

\title{Beyond Classical Instability Limits of Anisotropic Self-gravitating Fluid Configurations in Hu-Sawicki Inspired $f(R)$ Gravity}

\author{M. Yousaf}
\email{myousaf.math@gmail.com}
\affiliation{Department of Mathematics, Virtual University of Pakistan,\\ 54-Lawrence Road, Lahore 54000, Pakistan.}

\author{A. Rehman}
\email{attiqwatto786@gmail.com; atteeq.math@gmail.com}
\affiliation{Department of Mathematics, University of Management and
Technology,\\ Johar Town Campus, Lahore-54000, Pakistan.}

\author{M. Zeeshan Gul}
\email{mzeeshangul.math@gmail.com}
\affiliation{Tongji University, Shanghai 201804, China}

\author{Mohammed Zakarya}
\email{mzibrahim@kku.edu.sa}
\affiliation{King Khalid University, College of Science, Department of Mathematics,\\ P.O. Box 9004, 61413, Abha, Saudi Arabia.}

\author{Nadiah Zafer Al-Shehri}
\email{nadia@kku.edu.sa}
\affiliation{King Khalid University, College of Science, Department of Mathematics,\\ P.O. Box 9004, 61413, Abha, Saudi Arabia.}

\author{Imtiaz Khan}
\email{ikhanphys1993@gmail.com}
\affiliation{Department of Physics, Zhejiang Normal University, Jinhua, Zhejiang 321004, China}

\begin{abstract}
{In this draft, we investigate the dynamical instability of a restricted class of non-static, axially symmetric, self-gravitating fluid configurations within a Hu-Sawicki inspired $f(R)$ gravity model. The matter source is described by an anisotropic energy-momentum tensor containing three principal stresses and an off-diagonal stress component. For the adopted vorticity free geometry, conservation equations are formulated, and a linear perturbation scheme is applied to separate the equilibrium and time-dependent sectors. This procedure yields a collapse equation that governs the evolution of the perturbed compact configuration. The associated instability conditions are then derived in terms of the adiabatic index $\Gamma$ under the Newtonian and post-Newtonian approximations, whereas the resulting bounds show that the onset of instability depends not only on the stiffness of the fluid, but also on the background energy density, directional pressure anisotropies, metric perturbations, and higher-curvature contributions generated by the Hu-Sawicki model. The general relativistic limit is recovered by suppressing the modified gravity parameters, while the isotropic limit reproduces the classical Chandrasekhar threshold. These results demonstrate that curvature corrections and anisotropic stresses can appreciably modify the conventional instability conditions of axially symmetric compact systems.}
\end{abstract}

\maketitle

\section{Introduction}

The stability of relativistic compact formations, including spherical, cylindrical, and axially symmetric systems, is governed by a complex interplay among the matter distribution, fluid dynamics, density stratification, as well as the underlying gravitational scenario. A consistent treatment of these interconnected factors is therefore essential for determining the conditions under which a self-gravitating entity remains stable or undergoes collapse. Chandrasekhar established the limiting mass of white dwarfs as well as subsequently developed the dynamical stability criterion for relativistic stellar configurations as presented in \cite{chandrasekhar1957introduction,chandrasekhar1964dynamical}. Later, Herrera et al. extended this classical scenario to collapsing systems containing anisotropic and more general fluid distributions, showing that the corresponding stability limits may depart from the conventional Chandrasekhar bound \cite{herrera2009expansion,herrera2010collapsing,herrera2012dynamical,bamba2014cosmology,herrera2018tilted,herrera2022non,herrera2023expansion}. In the standard description, the adiabatic index $\Gamma=4/3$ serves as a reference threshold, with configurations generally tending toward instability for $\Gamma<4/3$ and stability for $\Gamma>4/3$. Nevertheless, anisotropy, spacetime geometry, as well as additional gravitational corrections can modify this critical value in both the Newtonian (N) and post-Newtonian (pN) regimes, with possible consequences for the final stages of gravitational collapse \cite{torres2005some,mitra2006gravitational,ivanov2010importance}.
The dynamical behavior of anisotropic compact systems also explored in a variety of modified gravity scenarios, where non-Einsteinian corrections were found to influence the equilibrium conditions and the associated instability ranges \cite{sharif2015instability,abbas2019dynamics,bhatti2023dynamical,yousaf2026effects,Almutairi2024impact} which emphasize the model dependent nature of gravitational stability as well as motivate further investigation in viable curvature based theories, whereas dynamical evolution of axially symmetric anisotropic sources by using perturbation approach with modified gravity studied in \cite{zubair2015evolution,yousaf2026analysis}. Some astrophysical investigations employed observational as well as computational approaches to study a broad range of energetic phenomena 
which emphasize the continuing importance of reliable theoretical scenarios for interpreting the dynamical behavior of astrophysical systems \cite{shah2019stability,bhatti2023stability,yousaf2026identifying,guo2026analyzing}.

{General relativity (GR) provides description for distinct gravitating phenomenon and the structural features of highly dense matter distributions. But, various observational and analytical issues assert that GR might not offer comprehensive interpretation for gravity at all scales. Specifically, the late-time cosmic acceleration suggested from Type Ia supernova findings is not properly described in GR without the incorporation of exotic essence of dark components \cite{riess1998observational,perlmutter1999measurements,riess2007new}. The $f(R)$ gravity is considered as the most basic and generally studied theory from all suggested modified formalisms of GR. In this case, the linear form of Ricci scalar $R$ is substituted with the function $f(R)$ in Einstein-Hilbert action which allows for more degrees of freedom that can play a role in the interpretation of rapid development of our cosmos without being dependent on exotic forms of energy and matter. A further compelling motivation for contemplating $f(R)$ theory emerges form quantified gravity and high-energy physics. The correction terms including higher order curvature variables are implicitly identified in generally considered actions determined from quantum field theory, string theory as well as several other methodologies related to quantum gravity. So, $f(R)$ theory can be regarded as a comparatively weak-energy representation of generalized gravitational theories. Moreover, $f(R)$ theory offers a comprehensive formalism for studying dynamical characteristics of compact objects. The added curvature terms modifies the gravity interactions potentially impacting the inner composition, balanced matter distribution, and tangible features of such systems. Subsequently, it can be asserted that the analysis of compact structures within the formalism of $f(R)$ gravity provides the pathway to evaluate the viability of alternative gravity theories employing astronomical results. It is important to mention that the stability analysis in the context of $f(R)$ gravity becomes more significant through the incorporation of extra curvature terms that results the modified form of gravity relations and additional degrees of freedom. These alterations can significantly contributes in determining the equilibrium state between attractive force of gravity and outward pressure affecting the dynamical stability of compact structures beyond to those suggested by GR. So, analysis of stability in $f(R)$ gravity validates the feasibility of determined solutions as well as offers better interpretation of the impacts caused by the alternative gravity in higher-density contexts.
The motivation for considering the Hu-Sawicki form of \(f(R)\) gravity is twofold. First, this model is one of the well-established viable \(f(R)\) models capable of describing late time cosmic acceleration without introducing an explicit dark energy component, while also allowing consistency with local gravitational constraints, secondly, the rational functional form of the Hu-Sawicki model introduces controlled higher curvature corrections through a small number of parameters, which makes it particularly suitable for studying how deviations from GR influence the internal equilibrium and collapse behavior of compact astrophysical entities. In the present context, these curvature corrections appear as effective geometric source terms or effective dark source terms in the modified field equations and consequently, directly affect the gravitational coupling, pressure gradients and perturbative evolution of the compact fluid distribution. Therefore, this model provides a physically motivated as well as mathematically tractable scenario for examining whether modified gravity corrections can shift the classical instability limits of compact self-gravitating systems. This makes the Hu-Sawicki model especially appropriate for studying how modified gravity influences the equilibrium and collapse behavior of high-density anisotropic astrophysical compositions.}

{The main aim of the present work is to determine the dynamical instability constraints for a restricted class of non-static, axially symmetric, anisotropic self-gravitating fluid configurations in the scenario of Hu-Sawicki inspired \(f(R)\) gravity. In particular, by applying a linear perturbation scheme to the modified field and conservation equations, we have obtained the collapse equation and extracted the corresponding instability criteria in terms of the adiabatic index \(\Gamma\) under both N and pN approximations. These results show how the usual Chandrasekhar type stability condition is modified by the combined effects of pressure anisotropy, axial symmetry, radial density inhomogeneity as well as higher curvature contributions coming from the Hu-Sawicki model. Consequently, the main outcome of our work is not merely the construction of the field equations, but the identification of the modified stability bounds that determine when an anisotropic compact configuration remains stable or enters the collapse regime in this extended gravitational background. In this sense, our analysis goes beyond the standard General Relativistic treatment by showing that the instability threshold is not governed only by the matter stiffness parameter \(\Gamma\), but also by the additional curvature contributions generated by the selected Hu-Sawicki \(f(R)\) model. The resulting stability bounds demonstrate that pressure anisotropy and higher-order curvature effects can shift the classical collapse condition and may enlarge the stability window for suitable choices of the model parameters.}
Modified gravity theories provide additional curvature contributions that can alter the equilibrium and dynamical stability of compact self-gravitating configurations. Related perturbative investigations in other modified-gravity frameworks have shown that matter anisotropy and additional geometric terms can appreciably shift the conventional instability limits.
Several investigations within modified gravity demonstrated that additional geometric contributions, pressure anisotropy, and matter geometry interactions can considerably alter the equilibrium and dynamical stability of compact self-gravitating configurations \cite{bhatti2023novel,yousaf2024unstable,naseer2024role,
yousaf2025viscous,rehman2025interpretation,asad2024evolution,naseer2025stellar,yousaf2025impact,gul2025impact,asad2025traversable}. These findings motivate a systematic examination of instability bounds beyond GR.

The present investigation is devoted to the onset of dynamical instability in a non-static, axially symmetric distribution of self-gravitating anisotropic matter, whereas departures from pressure isotropy are expected in highly compact astrophysical environments, where density gradients, strong interactions, magnetic fields, or directional stresses may cause the principal pressures to become unequal. Consequently, an anisotropic description can provide a broader representation of the internal dynamics of dense stellar systems than the conventional perfect fluid approximation. To determine how nonlinear curvature effects modify the collapse process, we work within a viable extension of GR based on the Hu-Sawicki prescription \cite{hu2007models,capozziello2008cosmography}. Modified versions of this curvature model also employed in astrophysical studies involving compact configurations and dark-matter supported structures studied in \cite{mustafa2022possibility}. The gravitational Lagrangian considered here is specified by
$f(R)
=
R
-
R_{c}
\frac{\Phi \left\{R/R_{c}\right\}^{n}}
{1+\Upsilon \left\{R/R_{c}\right\}^{n}},
$
where $R_{c}$ defines a characteristic curvature scale, while $\Phi$, $\Upsilon$, and $n$ are dimensionless constants controlling the deviation from the Einsteinian form, whereas the second term introduces nonlinear curvature effects that contribute to the effective gravitational source and may consequently alter the response of the fluid configuration to perturbations. It is useful to absorb the scale $R_{c}$ into redefined model parameters. Introducing
$
\breve{\Phi}
=
\Phi R_{c}^{\,1-n},
~
\breve{\Upsilon}
=
\Upsilon R_{c}^{-n},
$
the same function may be recast as
$
f(R)
=
R
-
\frac{\breve{\Phi}R^{n}}
{1+\breve{\Upsilon}R^{n}}.
$
This representation removes the explicit appearance of $R_{c}$ as well as expresses the curvature sector entirely through the effective combinations $\breve{\Phi}$ as well as $\breve{\Upsilon}$, while such a parametrization is more convenient for the perturbative treatment adopted below and does not modify the essential nonlinear character of the model.
Our analysis aims to establish the conditions under which the combined action of pressure anisotropy and Hu-Sawicki curvature corrections drives the axially symmetric fluid away from equilibrium. For this purpose, conservation laws and Ricci scalar are perturbed around a static background, and the resulting collapse equation is used to determine the admissible ranges of the adiabatic index in the N and pN limits. The paper is arranged as follows. Section~II presents the basic generic field equations associated with the chosen $f(R)$ model. Section~III develops the corresponding conservation relations and provides the curvature scalar for the restricted axially symmetric spacetime. The perturbation formalism, together with the derivation of the collapse equation, is discussed in Section~IV. The N and pN instability conditions are examined in Section~V, while Section~VI contains the main conclusions, whereas the lengthy curvature source quantities are listed in the Appendix.

\section{Axially Symmetric Geometry and Field Equations:}

{The feasibility of any modified theory of gravity is determined by its compatibility with physical scales, ranging from the primordial development of universe to the dynamics of extremely dense astronomical structures \cite{buchdahl1970non,nojiri2005modified,nojiri2011unified,olmo2011palatini,capozziello2011extended,harko2011f,sharif2016energy,
lovelock1971einstein}. In the formalism of $f(R)$ theory, the identical variations pertaining to the Einstein-Hilbert action affecting the behaviour of cosmos on larger scale also have implications for gravitating field equations regulating local astronomical objects. As a result, the extra degrees of freedom provided to address cosmic acceleration have the potential to modify the inner configurations of celestial objects, equilibrium constraints, and dynamical stability. The relationship between cosmological and astrophysical domains is emerged from the fact that both are determined by the similar fundamental gravitational force. Considering that the curvature scales are substantially different, the modifications retained in $f(R)$ function are expressed as the corrections to gravitating dynamics across both low-curvature cosmic scenarios and high-curvature inner structures. So, it can be asserted that compact structures can be employed as physical laboratories for evaluating the feasibility of alternative gravity theories in the context of strong gravitating fields along with the cosmological data that analyze gravity on larger scales. Moreover, an effective formalism of alternative gravity should comply with both cosmological restraints as well as the astrophysical constraints. A mathematical framework interpreting the late-time cosmic acceleration but incapable of establishing stable and dynamically viable compact systems will have to confront certain issues related to its feasibility. Therefore, the stability analysis related to compact structures with in the context of $f(R)$ theory serves as an important connection between cosmic mechanisms and astrophysical estimations, providing a comprehensive formalism for evaluating gravity at various length and energy scales.}
The action governing the Hu-Sawicki inspired $f(R)$ gravitational model is written as
\begin{equation}\label{1}
I=\frac{1}{16\pi G}\int\left\{R-\dfrac{\breve{\Phi} R^{n}}{1+\breve{\Upsilon} R^{n}}\right\}d^4x\sqrt{-g} +I_{m}~.
\end{equation}
Here $g$ is trace of metric tensor and $G$ serves as the gravitational constant, while $I_{m}$ is matter action. The nonlinear dependence on the Ricci scalar introduces additional curvature contributions that modify the effective gravitational dynamics. These corrections are relevant to the equilibrium and stability analysis of the compact fluid configuration considered here. Generic field equations are derived after the metric variation of Eq. \eqref{1} as

\begin{equation}\label{2a}
G_{\chi\zeta}=T_{\chi\zeta}^{(eff)}.
\end{equation}
here, $G_{\varrho \zeta}=R_{\varrho \zeta}-\frac{1}{2}Rg_{\varrho \zeta}$ and $T_{\varrho \zeta}^{(eff)}$ is  classified as
\begin{align}\nonumber
T_{\chi\zeta}^{(eff)}&=\frac{1}{1 - \frac{n\breve{\Phi} R^{n-1}}{\left\{1 + \breve{\Upsilon} R^n\right\}^2}}\left\{T_{\chi\zeta}^{(m)}+\frac{1}{2}\left\{R - \frac{\breve{\Phi} R^n}{1 + \breve{\Upsilon} R^n}-R\left\{1 - \frac{n\breve{\Phi} R^{n-1}}{\left\{1 + \breve{\Upsilon} R^n\right\}^2}\right\}\right\}g_{\chi\zeta}\right.
\\\label{2a}&\left.+(\nabla_{\chi}\nabla_{\zeta}-g_{\chi\zeta}\Box)\left\{1 - \frac{n\breve{\Phi} R^{n-1}}{\left\{1 + \breve{\Upsilon} R^n\right\}^2}\right\}\right\},
\end{align}
In modified field equations $\nabla_{\varrho}$ specifies the covariant derivative. The specific representation of $T$ through the incorporation of explicit stresses and pressure is important  in obtaining the solution of Einstein's field equations associated with the anisotropic axially symmetric fluid dispersion, and leads to the comprehensive description of compact objects. We consider the subsequent representation of energy-momentum tensor
\begin{eqnarray}\label{3a}
T_{\chi\zeta}^{(m)} &=& (\varrho+P)V_{\chi}V_{\zeta} + Pg_{\chi\zeta} + \Pi_{\chi\zeta},
\end{eqnarray}
here, energy density, surface pressure, velocity four-vector, and the anisotropic stress tensor are denoted by $\rho$, $P$, $v_{\chi}$, and $\Pi_{\varrho \zeta}$ respectively, while, this mathematical representation incorporates the effects of surface pressure in different directions. The mathematical expression for $\Pi_{\varrho \zeta}$ is
\begin{equation}\label{a3a}
\Pi_{\chi\zeta} = \left(\mathcal{K}_{\chi}\mathcal{K}_{\zeta} - \frac{h_{\chi\zeta}}{3}\right)(P_{xx} - P_{zz})+ 2P_{xy}\mathcal{K}_{(\chi}L_{\zeta)}
- (P_{zz}-P_{yy})\left(L_{\chi}L_{\zeta} - \frac{h_{\chi\zeta}}{3}\right)
,
\end{equation}
in which $P_{xx}$, $P_{yy}$, and $P_{zz}$ denote the components of surface pressure in respective directions, and $P_{xy}$ represents the shear stress, whereas the term $K_{\chi}$ and $L_{\chi}$ specify the unit four vectors. The symmetrical tensor is expressed as
\begin{equation}
\mathcal{K}_{(\chi}L_{\zeta)} = \frac{1}{2} \left\{ \mathcal{K}_{\chi}L_{\zeta} + \mathcal{K}_{\zeta}L_{\chi} \right\}
\end{equation}
The pressure can be described by means of surface pressures in following form
$
P = \frac{1}{3}(P_{xx} + P_{yy} + P_{zz}).
$
The preceding average serves as a scalar study of internal forces that takes into account the pressure distributions in the x, y, and z directions. The projection tensor is described in following form
$
h_{\chi\zeta} = g_{\chi\zeta} + V_{\chi}V_{\zeta},
$
where, $g_{\chi\zeta}$ specifies the spacetime metric, while, the projection tensor provides a distribution of values corresponding to the four velocity, effectively separating through space and time components in the interrelated framework, and the surface pressure in addition to the shear stress, are considered as primary components of the stress tensor which comply with the relationship $P_{xy}=P_{yx}$ establishing the symmetry, and the restraints  $P_{xx}\neq P_{yy}\neq P_{zz}$ demonstrates the anisotropic distribution, that corresponds to the spatial restraints in the stress distribution of matter.

The line element is a mathematical formalism which is used to determine the distance and time within curved or flat geometry, i.e., the spacing between two adjacent spatial phenomenon, that provides description for the bending and stretching of spacetime caused by substantial variables including gravity, mass, or energy, that primarily determine the configuration of cosmos on local and galatic scales.

Generally, it is described as:
$ds^{2} = g_{\chi\zeta} d x^\chi d x^\zeta,$ where \( g_{\chi\zeta} \) refers to the various parts of metric tensor whereas, \( d x^\chi\) are infinitesimal coordinate displacements.
The axial geometry provides appropriate context for the analysis of compact structures, accelerated matter compositions, and gravitational collapse other than spherical symmetry, whereas symmetric spacetimes are contemplated as the generalized form of spherical geometry. This encompasses a wider range of cosmological compact structures with rotation, distortion, directionally oriented anisotropy, whereas axial symmetry within the formalism of alternative theory allows for the development of novel coupling-driven stresses and directional impacts. In this regard it provides the comprehensive interpretation for the impacts implied by the matter geometry relationships on stability and dynamical development of self-gravitational highly dense matter compositions. To analyze the structural characteristics of compact structures, we contemplate axially symmetric compositions, such systems including BHs and neutron stars, which possess distinctive space-time characteristics. The generalized form of axial line element is expressed in subsequent form:
\begin{equation}\label{1aa}
\centering
\left.\begin{split}
d s^2=&{\Sigma}_{\chi, \zeta=t, r, \Theta, \varphi} g_{\chi\zeta} d x^\chi d x^\zeta=-A^2(t,r,\Theta)dt^{2}+B^{2}(t,r,\Theta)
\Sigma_{\rho, \upsilon=r, \Theta} \bar{g}_{\rho \upsilon} d x^\rho d x^\upsilon+ 2Y(t,r,\Theta) dt d\Theta
\\& + 2H(t,r,\Theta) dt d\varphi+C^2(t,r,\Theta)d\varphi^2,~~\text{where}~~
\Sigma_{\rho, \upsilon=r, \Theta} \bar{g}_{\rho \upsilon} d x^\rho d x^\upsilon=dr^2+r^2d\Theta^2.
\end{split}~~\right\}......
\end{equation}
here, A, B, C, Y, and H are metric coefficients, which have dependence on time coordinate whereas $r$ and $\Theta$ represent the spacial coordinates, further they goven the interplay of spacetime. Thus, $A^2$ signifies the temporal part, $B^2$ defines the angular and radial axes, and $C^2$ is related to the axial coordinate while, Y, H offer non-diagonal conceptions that demonstrate time and angular directions. The appearance of non-diagonal terms, Y and H in metric tensor yields the derivation of non-linear equations and these terms specify rotation and interaction between spatial and tangential parts, rendering our analysis challenging. It is generally considered that non-static axial geometry includes rotational and reflectional effects derived from non-diagnol metric factors. Several approaches have focused on a less complicated class of axial geometry to mathematically address the unstable constraints related to axially symmetric spacetime. Thus, axially symmetric metric describing the effects of reflection is written as
\begin{equation}\label{1aaa}
\centering
\left.\begin{split}
d s^2=&\Sigma_{\chi, \zeta=t, r, \Theta, \varphi} g_{\chi\zeta} d x^\chi d x^\zeta=-A^2(t,r,\Theta)dt^{2}+B^{2}(t,r,\Theta)
\Sigma_{\rho, \upsilon=r, \Theta} \bar{g}_{\rho \upsilon} d x^\rho d x^\upsilon+ 2Y(t,r,\Theta) dt d\Theta
\\& +C^2(t,r,\Theta)d\varphi^2,~~~
\Sigma_{\rho, \upsilon=r, \Theta} \bar{g}_{\rho \upsilon} d x^\rho d x^\upsilon=dr^2+r^2d\Theta^2.
\end{split}~~\right\}......
\end{equation}
The analysis of compositions involving non-diagonal metric  parts is complicated, and the basic Einstein equations of motion render it much more difficult to acquire precise analytical solutions. A vorticity-free approximation is generally considered to overcome these challenges. The assumption of absence of circular motion, reflection impacts, or rotation along symmetrical axis can significantly reduce the combination of metric components  $dtd\Theta$ and $dtd\varphi$. The consideration of vorticity-free significantly reduces the complexity of the system, resulting the simplified form of field equations, so, we constrain our focus to a particular type of non-static, axially symmetric, collapsed highly dense matter compositions. In this context, meridional and rotational motions corresponding to symmetric axis are ignored. The related modified line element, comprising three independent metric functions has the subsequent representation
\begin{equation}\label{1a}
\centering
\left.\begin{split}
d s^2=&\Sigma_{\chi, \zeta=t, r, \Theta, \varphi} g_{\chi\zeta} d x^\chi d x^\zeta=-A^2(t,r,\Theta)dt^{2}+B^{2}(t,r,\Theta)
\Sigma_{\rho, \upsilon=r, \Theta} \bar{g}_{\rho \upsilon} d x^\rho d x^\upsilon+C^2(t,r,\Theta)d\varphi^2,
\\&
\Sigma_{\rho, \upsilon=r, \Theta} \bar{g}_{\rho \upsilon} d x^\rho d x^\upsilon=dr^2+r^2d\Theta^2.
\end{split}~~\right\}......
\end{equation}
here, $A$, $B,$ and $C$ are contemplated as positive functions which implies that the axially symmetric metric retains its fundamental attributes. This representation of metric is generally considered in modelling of compact structure, demonstrating its relevance in the understanding of related configuration and physical dynamics. In the framework of zero vorticity, simplest representation of mathematical formulas can be derived that emphasize crucial features of physical structure allowing the thorough interpretation of contemplated compact structures.
Both $K_{\chi}$ and $L_{\chi}$ have the subsequent mathematical representations:
\begin{equation}
\mathcal{K}_{\chi} = \delta^{1}_{\chi}Z, \quad L_{\chi} = r\delta^{2}_{\chi}Z.
\end{equation}
These vectors establish a standard framework for evaluating the linear stress components in interrelated structures. The velocity four vector is expressed as
\begin{equation}
V_{\chi} = -\delta^o_{\chi}X,
\end{equation}
where, $\delta^o_{\chi}$ specifies the dominated temporal component, and $X$ relates to the normalization and this vector determines  the behaviour of matter in the internal composition of the spacetime. The interaction between  $K_{\chi}$, $v_{\chi}$, and the $L_{\chi}$ establishes a comprehensive framework which permits the decomposition of significant quantities into components which are consistent with the fundamental directions of the system, leading to the detailed description for anisotropic pressure ans stress. Moreover, anisotropic stresses have key relevance in modelling compact systems, and it is important to mention that evolution and stability of compact structures are significantly impacted by these stresses. The relationship between $K_{\chi}$, $v_{\chi}$, and the $L_{\chi}$ are particularly useful in co-moving frames because they develop a direct interaction with the geometrical configuration of these systems. As a result, the energy momentum tensor associated with the usual matter allows the detailed framework  for interpreting axially symmetric configurations and corresponding parameters, as specified by directional vectors and metric relationships, providing a comprehensive description for the relationship between isotropic pressure, energy density, anisotropic stresses and spacetime. Furthermore, this formalism has important implications for determining stability and basic dynamics of compact compositions of matter.
One can find the non-zero components of the modified Einstein tensor with $f(R)$ corrections in \cite{yousaf2025dynamical}.

\section{Conservation Equations and Ricci Scalar:}

{The transition of a self-gravitating configuration from a state of hydrostatic equilibrium to gravitational collapse must remain compatible with the fundamental physical principles governing the system, particularly the conservation of energy and momentum. In the present analysis, these conservation laws are investigated within the scenario of the Hu-Sawicki $f(R)$ gravity model. As an extension of GR, $f(R)$ gravity provides an appropriate setting for examining the dynamical evolution of an axially symmetric compact entity as it departs from equilibrium and enters a collapsing regime. For a relativistic matter distribution, the local conservation of energy and momentum is expressed as
\begin{equation}\nonumber
\nabla_{\zeta}T^{\zeta\chi}=0,
\end{equation}
where $\nabla_{\zeta}$ denotes the covariant derivative and $T^{\zeta\chi}$ represents the energy-momentum tensor of the matter source. We consider a restricted class of non-static and axially symmetric collapsing fluid configurations characterized by three independent metric functions, as introduced in Eq.~\eqref{1a}, together with the energy-momentum tensor of ordinary matter given in Eqs.~\eqref{3a} and \eqref{a3a} together. Whereas the independent conservation equations are obtained by assigning appropriate values to the free index $\chi$, in particular, the choices $\chi=0=t$, $\chi=1=r$ and $\chi=2=\Theta$ yield the temporal, radial, and angular components of the conservation law, respectively. The remaining repeated index $\zeta$ is summed over all spacetime coordinates in accordance with the Einstein summation convention. Throughout the analysis, an overdot denotes differentiation with respect to the temporal coordinate $t$, whereas a prime represents differentiation with respect to the radial coordinate $r$, while differentiation with respect to the angular coordinate $\Theta$ is written explicitly. These derivatives describe the temporal evolution, radial variation, and angular dependence of the physical and geometrical variables involved in the collapsing configuration. The temporal, radial, and angular components collectively provide a complete description of the local exchange as well as redistribution of energy and momentum within the axially symmetric fluid, consequently, the conservation equations ensure that the dynamical evolution of the system, whether it remains close to equilibrium or proceeds toward collapse, is consistent with the underlying conservation principles in considered Hu-Sawicki inspired $f(R)$ gravity. For the adopted axially symmetric geometry, the conservation law generates three independent dynamical equations governing the balance of energy and momentum.}

\textbf{{Energy Conservation Equation $(\chi=0=t)$}}

{The temporal component is obtained by setting $\chi=0=t$ and may be identified as the energy conservation equation, while it describes the evolution of the matter energy density together with the contributions arising from stresses, fluid expansion, and energy transport. Its explicit form is given by}

\begin{align}\nonumber
&-\frac{1}{r^{2}}\left\{\frac{W_{02}}{B^{2}A^{2}(1-\frac{\breve{\Phi} n R^{\,n-1}}{(1+\breve{\Upsilon} R^{n})^{2}})}\right\}^{\Theta}
-\left\{\frac{W_{01}}{B^{2}A^{2}(1-\frac{\breve{\Phi} n R^{\,n-1}}{(1+\breve{\Upsilon} R^{n})^{2}})}\right\}'+\left[\dot{\varrho}+\left\{\left\{2\frac{\dot{B}}{B}+\frac{\dot{C}}{C}\right\}-{\left\{1-\frac{\breve{\Phi} n R^{\,n-1}}{(1+\breve{\Upsilon} R^{n})^{2}} \right\}^{.}}\right\}\varrho
\right.\\\nonumber&\left.+\left\{\frac{\dot{C}}{C}P_{zz}+\frac{\dot{B}}{B}P_{yy}\right\}
+\left\{\frac{(R-\dfrac{\breve{\Phi} R^{n}}{1+\breve{\Upsilon} R^{n}})\left\{1-\frac{\breve{\Phi} n R^{\,n-1}}{(1+\breve{\Upsilon} R^{n})^{2}} \right\}^{.}}{2\left\{1-\frac{\breve{\Phi} n R^{\,n-1}}{(1+\breve{\Upsilon} R^{n})^{2}} \right\}}-\frac{1}{2}\left\{ R-\dfrac{\breve{\Phi} R^{n}}{1+\breve{\Upsilon} R^{n}} \right\}^{.}+\frac{\dot{B}R(1-\frac{\breve{\Phi} n R^{\,n-1}}{(1+\breve{\Upsilon} R^{n})^{2}})}{BA}\right.\right.\\\nonumber &\left.
+\frac{1}{2}\dot{R}(1-\frac{\breve{\Phi} n R^{\,n-1}}{(1+\breve{\Upsilon} R^{n})^{2}})\right\}
+\frac{C^{\Theta}}{C}W_{33}+2\left\{\frac{\dot{B}}{B}+\frac{\dot{A}}{A}
+\frac{\dot{C}}{2C}\right\}W_{00}-\frac{1}{B^{2}}\left\{\frac{3A'}
{A}+\frac{1}{r}+\frac{2B'}{B}+\frac{C'}{C}\right\}W_{01}+\frac{B^{\Theta}}{B}W_{11}
\\\label{10a} &
\left.+\frac{B^{\Theta}}{B}W_{22}-\frac{1}{r^{2}B^{2}}\left\{\frac{3A^{\Theta}}{A}+\frac{C^{\Theta}}{C}+
\frac{2B^{\Theta}}{B}\right\}W_{02}\right]\frac{1}{A^{2}\left\{1-\frac{\breve{\Phi} n R^{\,n-1}}{(1+\breve{\Upsilon} R^{n})^{2}} \right\}}+\left\{\frac{W_{00}}{A^{2}(1-\frac{\breve{\Phi} n R^{\,n-1}}{(1+\breve{\Upsilon} R^{n})^{2}})}\right\}^{.}=0.
\end{align}

{This relation guarantees the local conservation of energy during its transport and redistribution throughout the collapsing fluid configuration.}

\textbf{{Radial Momentum Conservation Equation $(\chi=1=r)$}}

{The second conservation equation is obtained by selecting the radial component, $\chi=1=r$, while it governs the balance of forces acting along the radial direction and describes how the radial momentum of the fluid evolves under the combined influence of pressure gradients, anisotropic stresses, gravitational effects, and the geometrical properties of the spacetime. The resulting radial momentum conservation equation can be written in the following form:}

\begin{align}\nonumber
&\left\{\frac{W_{11}}{B^{2}(1-\frac{\breve{\Phi} n R^{\,n-1}}{(1+\breve{\Upsilon} R^{n})^{2}})}\right\}^{'}+\left\{\frac{W_{12}}{B^{2}(1-\frac{\breve{\Phi} n R^{\,n-1}}{(1+\breve{\Upsilon} R^{n})^{2}})}\right\}^{\Theta}+\frac{1}{B^{2}(1-\frac{\breve{\Phi} n R^{\,n-1}}{(1+\breve{\Upsilon} R^{n})^{2}})}\left[\frac{A'\varrho}{A}\right.+\left\{\frac{1}{r}+\frac{C}{C}+\frac{B}{B}
-\frac{(1-\frac{\breve{\Phi} n R^{\,n-1}}{(1+\breve{\Upsilon} R^{n})^{2}})'}{1-\frac{\breve{\Phi} n R^{\,n-1}}{(1+\breve{\Upsilon} R^{n})^{2}}}\right\}P_{xx}
\\\nonumber &+P'_{xx}+\frac{P_{xy}^{\Theta}}{r}-\frac{C^{'}}{C}P_{zz}-\left\{\frac{1}{r}+\frac{B'}{B}\right\}P_{yy}
+\frac{1}{r}\left\{\left\{\frac{2B^{\Theta}}{B}+\frac{C^{\Theta}}{C}-\frac{(1-\frac{\breve{\Phi} n R^{\,n-1}}{(1+\breve{\Upsilon} R^{n})^{2}})^{\Theta}}{1-\frac{\breve{\Phi} n R^{\,n-1}}{(1+\breve{\Upsilon} R^{n})^{2}}}+\frac{A^{\Theta}}{A}\right\}
+\left\{R-\dfrac{\breve{\Phi} R^{n}}{1+\breve{\Upsilon} R^{n}}\right\}^{\Theta}\right\}P_{xy}\\\nonumber
&\left.\frac{1}{2}(R-\dfrac{\breve{\Phi} R^{n}}{1+\breve{\Upsilon} R^{n}})^{'}-\frac{\left\{ R-\dfrac{\breve{\Phi} R^{n}}{1+\breve{\Upsilon} R^{n}} \right\}^{'}(R-\dfrac{\breve{\Phi} R^{n}}{1+\breve{\Upsilon} R^{n}})}{2(1-\frac{\breve{\Phi} n R^{\,n-1}}{(1+\breve{\Upsilon} R^{n})^{2}})}
-\frac{1}{2}R'(1-\frac{\breve{\Phi} n R^{\,n-1}}{(1+\breve{\Upsilon} R^{n})^{2}})\right]-{\left\{\frac{W_{01}}{B^{2}A^{2}(1-\frac{\breve{\Phi} n R^{\,n-1}}{(1+\breve{\Upsilon} R^{n})^{2}})}\right\}^{.}}
\\\nonumber&+\frac{1}{B^{2}(1-\frac{\breve{\Phi} n R^{\,n-1}}{(1+\breve{\Upsilon} R^{n})^{2}})}\left\{\left\{\frac{3B'}{B}+\frac{1}{r}+\frac{A'}{A}
+\frac{C'}{C}\right\}W_{11}+\frac{A'}{A}W_{00}-\frac{1}{A^{2}}\left\{\frac{\dot{C}}{C}+\frac{4\dot{B}}{B}
+\frac{\dot{A}}{A}\right\}W_{01}+\left\{\frac{4B^{\Theta}}{B}+\frac{C^{\Theta}}{C}+\frac{A^{\Theta}}{A}\right\}W_{12}
\right.
\\\label{11a}&\left.
-\left\{\frac{1}{r}+\frac{B'}{B}\right\}W_{22}-\frac{C'}{2C}W_{33}\right\}=0
\end{align}

{This relation provides the relativistic counterpart of the hydrostatic equilibrium condition in the radial direction, whereas the inclusion of the anisotropic stress tensor, $\Pi_{\zeta\chi}$, introduces additional force contributions and consequently alters the equilibrium and stability properties of the compact configuration.}

\textbf{{Angular Momentum Conservation Equation $(\chi=2=\Theta)$}}

{For an anisotropic matter distribution, the conservation law also possesses an independent component along the angular direction, whereas this equation is obtained by fixing the free index as $\chi=3=\Theta$ and describes the balance of momentum and stresses associated with angular variations in the axially symmetric fluid. In the adopted coordinate convention, the angular component is derived by selecting the index corresponding to the $\Theta$ coordinate, however the resulting conservation equation can be expressed in the following form:}

\begin{align}\nonumber
&\left[\left\{1-\frac{\left\{ R-\dfrac{\breve{\Phi} R^{n}}{1+\breve{\Upsilon} R^{n}} \right\}^{\Theta}}{1-\frac{\breve{\Phi} n R^{\,n-1}}{(1+\breve{\Upsilon} R^{n})^{2}}}
+\frac{A^{\Theta}}{A}\right\}\varrho+\left\{\left\{\frac{2rB'}{B}+\frac{rC'}{C}-\frac{r\left\{1-\frac{\breve{\Phi} n R^{\,n-1}}{(1+\breve{\Upsilon} R^{n})^{2}} \right\}^{'}}{1-\frac{\breve{\Phi} n R^{\,n-1}}{(1+\breve{\Upsilon} R^{n})^{2}}}+\frac{rA'}{A}\right\}\right\}P_{xy}-\frac{B^{\Theta}P_{xx}}{B}+P_{zz}\frac{C^{\Theta}}{C}\right.
\\\nonumber &+\left\{\frac{B^{\Theta}}{B}+\frac{A^{\Theta}}{A}-\frac{f^{\Theta}_{R}}{f_{R}}
+\frac{C^{\Theta}}{C}\right\}P_{yy}+\left.\frac{\left\{ R-\dfrac{\breve{\Phi} R^{n}}{1+\breve{\Upsilon} R^{n}} \right\}^{\Theta}}{2}-\frac{\left\{1-\frac{\breve{\Phi} n R^{\,n-1}}{(1+\breve{\Upsilon} R^{n})^{2}} \right\}^{\Theta}(R-\dfrac{\breve{\Phi} R^{n}}{1+\breve{\Upsilon} R^{n}})}{2(1-\frac{\breve{\Phi} n R^{\,n-1}}{(1+\breve{\Upsilon} R^{n})^{2}})}-\frac{1}{2}R^{\Theta}(1-\frac{\breve{\Phi} n R^{\,n-1}}{(1+\breve{\Upsilon} R^{n})^{2}})\right]
\\\nonumber&\times \frac{1}{r^{2}B^{2}(1-\frac{\breve{\Phi} n R^{\,n-1}}{(1+\breve{\Upsilon} R^{n})^{2}})}
-\frac{1}{r^{2}}\left\{\frac{W_{02}}{B^{2}A^{2}(1-\frac{\breve{\Phi} n R^{\,n-1}}{(1+\breve{\Upsilon} R^{n})^{2}})^2}\right\}^{.}+\left\{\frac{W_{12}}{rB^{2}(1-\frac{\breve{\Phi} n R^{\,n-1}}{(1+\breve{\Upsilon} R^{n})^{2}})}\right\}^{'}+\frac{1}{r^{2}}\left\{\frac{W_{22}}{B^{2}(1-\frac{\breve{\Phi} n R^{\,n-1}}{(1+\breve{\Upsilon} R^{n})^{2}})}\right\}^{\Theta}
\\\nonumber &+\frac{1}{r^{2}B^{2}}\left\{\frac{A^{\Theta}W_{00}}{A(1-\frac{\breve{\Phi} n R^{\,n-1}}{(1+\breve{\Upsilon} R^{n})^{2}})}-\frac{B^{\Theta}W_{11}}{B(1-\frac{\breve{\Phi} n R^{\,n-1}}{(1+\breve{\Upsilon} R^{n})^{2}})}
-\frac{C^{\Theta}W_{33}}{C(1-\frac{\breve{\Phi} n R^{\,n-1}}{(1+\breve{\Upsilon} R^{n})^{2}})}\right\}+\frac{1}{r^{2}B^{2}(1-\frac{\breve{\Phi} n R^{\,n-1}}{(1+\breve{\Upsilon} R^{n})^{2}})}\left[\left\{\frac{3B^{\Theta}}{B}+\frac{C^{\Theta}}{C}+\frac{A^{\Theta}}{A}\right\}\right.
\\\label{12a}&\times W_{22}\left.
-\frac{1}{A^{2}}\left\{\frac{4\dot{B}}{B}+\frac{\dot{C}}{C}+\frac{\dot{A}}{A}\right\}W_{02}\right]+\frac{4}{rB^{2}(1-\frac{\breve{\Phi} n R^{\,n-1}}{(1+\breve{\Upsilon} R^{n})^{2}})}\left\{\frac{B'}{B}+\frac{A'}{4A}+\frac{3}{4r}+\frac{C'}{4C}\right\}W_{12}+rP'_{xy}+P_{yy}^\Theta=0.
\end{align}

{This relation governs the transfer and redistribution of momentum along the angular direction. In these conservation expressions the additional geometric contributions originate from the curvature corrections associated with the adopted modified gravity model. Taken together, the temporal, radial, and angular conservation equations preserve the dynamical consistency of the axially symmetric gravitational configuration.}
Ricci scalar for our considered axially symmetric geometry given as,
\begin{align}
\\\nonumber&R=\frac{2}{B^{2}}\left[\left\{\dot{B}B+\frac{B^2\dot{C}}{C}\right\}\frac{\dot{A}}{A^{3}}+B'\left\{\frac{1}{Br}+\frac{B''}{BB'}-\frac{B'}{B^2}\right\}+A'\frac{1}{A}\left\{\frac{C'}{C}+\frac{A''}{A'}-\frac{1}{r}\right\}+\frac{B^{\Theta}}{r^2}\left\{\frac{B^{\Theta\Theta}}{B^{\Theta}}-\frac{B^{\Theta}}{B}\right\}
-\frac{B^2\ddot{C}}{A^{2}C}\right.\\\label{13a}&+\left\{\frac{1}{A}\left\{\frac{1}{r^2}A^{\Theta\Theta}+\frac{A^{\Theta}C^{\Theta}}{r^2C}\right.\right.
\left.\left.\left.+\frac{C^{\Theta}A}{r^2C}\right\}\right\}-\frac{2\dot{B}B^2}{A^2B}\left\{\frac{\dot{C}}
{C}+\frac{\dot{B}}{2B}+\frac{\ddot{B}}{\dot{B}}\right\}
+\frac{C'}{rC}\left\{\frac{C''}{C'}+1\right\}\right].
\end{align}

\section{Perturbation Approach}

The perturbation approach serves as mathematical framework for assessing physical instability in gravitational contexts as well as an explanation of the conduct of gravitational field through the extension of equations regulating the gravitational field with consideration of minor fluctuations or perturbations within the assumed foundational position. A systematic methdalogy is important for determining  stability as nonlinear formalisms generally employ complicated analytical solutions, whereas considering and validating appropriate perturbation approach has significant relevance in the comprehensive analysis affecting cosmological evaluations of compact structures.
After the contemplation of perturbation \cite{herrera2012dynamical} within the formalism of revised $f(R)$ gravity, in this section of paper, we are intended to analyze the steady dynamics of restricted non static axially symmetric self-gravitational matter compositions. We have studied that the set of substantially non-linear revised field equations remains devoid of generic solutions in this modified theory of gravity. It can be asserted that, while the composition of highly dense matter is mainly determined by a particular radial coordinate as well as it depends on time coordinate with passage of time that correspond to the static part of a particular physical parameter, is represented by the subscript zero.

The general perturbation approach related to Ricci scalar and metric coefficients is defined in subsequent form:
\begin{align}\label{14a}&
{\bigsqcup}_{\nu}(t,r,\Theta)={\bigsqcup}_{\nu0}(r,\Theta)+\epsilon \aleph_{\nu}(t)\digamma_{\nu}(r,\Theta)
\\\label{23a}&
{\bigvee}_{\sigma}(t,r,\Theta) = {\bigvee}_{\sigma0}(r,\Theta) + \epsilon \bar{\bigvee}_{\sigma}(t,r,\Theta),
~~\text{and}~~\varrho(t,r,\Theta)=\varrho_o(r,\Theta)+ \bar{\varrho}(t,r,\Theta)\epsilon.
\end{align}

{
In Eq.~\eqref{14a}, the index $\nu$ is introduced as a collective label for the perturbed geometrical quantities, namely the three metric functions and the Ricci scalar. More precisely,
\[
\nu=1,2,3,4,
\qquad
\bigsqcup_{1}=A,\quad
\bigsqcup_{2}=B,\quad
\bigsqcup_{3}=C,\quad
\bigsqcup_{4}=R,
\]
with the associated spatial perturbation profiles defined by
\[
\digamma_{1}=a,\quad
\digamma_{2}=b,\quad
\digamma_{3}=c,\quad
\digamma_{4}=e.
\]
Here, $A$, $B$, and $C$ denote the metric potentials, whereas $R$ represents the Ricci scalar. The functions $a$, $b$, $c$, and $e$ describe the corresponding spatial parts of their perturbations. In addition, the temporal dependence is assumed to be common to all perturbed geometrical variables, so that
\[
\aleph_{1}=\aleph_{2}=\aleph_{3}=\aleph_{4}=Y.
\]
This notation therefore provides a compact representation of the linear perturbation expansion for the complete set of geometrical quantities.

In Eq.~\eqref{23a}, the subscript \(\sigma\) is introduced separately to denote the different physical fluid variables components corresponding to various pressures and energy density with \(\sigma = 1, 2, 3, 4\) indexing these pressures and shear stresses. Here, \(\bigvee_{\sigma0}\) denotes the static equilibrium values of these fluid variables, and \(\bar{\bigvee}_{\sigma}\) represents their perturbations with explicit dependence on time, radius, and angle.
The reason behind using distinct subscripts \(\nu\) and \(\sigma\) is to clearly differentiate between geometric (metric and curvature) perturbations labeled by \(\nu\) and matter variables (pressure and density components) labeled by \(\sigma\). Briefly, here \( {\bigvee}_1 = P_{{xx}} \), \( {\bigvee}_2 = P_{{yy}} \), \( {\bigvee}_3 = P_{{zz}} \), \( {\bigvee}_4 = P_{{xy}}=P_{{yx}} \), also, \( {\bigvee}_{\sigma0}(r,\Theta) \) represent the non perturbed parts, while \( \bar{{\bigvee}}_{\sigma}(t,r,\Theta) \) introduces the perturbative components with time and angular dependencies which are stated as \( {\bigvee}_{10} = P_{{xx}0} \), \( {\bigvee}_{20} = P_{{yy}0} \), \( {\bigvee}_{30} = P_{{zz}0} \), \( {\bigvee}_{40} = P_{{xy}0}=P_{{yx}0}\), also \(\bar{{\bigvee}}_{1}=\bar{P}_{{xx}}\), \(\bar{{\bigvee}}_{2}=\bar{P}_{{yy}}\), \(\bar{{\bigvee}}_{3}=\bar{P}_{{zz}}\), and \(\bar{{\bigvee}}_{4}=\bar{P}_{{yx}}=\bar{P}_{{xy}}\). Here, \( \bigvee_o(r,\Theta) \) represents the static part, while \( \bar{\bigvee}(t,r,\Theta) \) are perturbative components introducing time and angular dependencies in this case, $1\gg\epsilon>0$ and we have perturbation scheme for considered model as 
\begin{align}\label{25a}
f(R)&=R_o-\frac{\breve{\Phi} R_o^{n}}{1+\breve{\Upsilon} R_o^{n}}+\epsilon\,Ye\left[1-\frac{\breve{\Phi} n R_o^{\,n-1}}{(1+\breve{\Upsilon} R_o^{n})^{2}}\right],
\end{align}

\subsection{Static Layout}

By applying the perturbation scheme to Eqs.~\eqref{10a} to \eqref{12a}, the static components of the conservation equations are obtained as follows:

\begin{align}\nonumber
&T^{0\zeta}_{;\zeta}=\left\{\frac{W_{00}}{A_o^{2}(1 - \frac{n \breve{\Phi} R_o^{\,n-1}}{(1+\breve{\Upsilon} R_o^{n})^{2}})}\right\}^{.}-\left\{\frac{W_{01}}{A_o^{2}B^{2}_o(1 - \frac{n \breve{\Phi} R_o^{\,n-1}}{(1+\breve{\Upsilon} R_o^{n})^{2}})}\right\}^{'}
-\frac{1}{r^2}\left\{\frac{W_{02}}{A^{2}_oB_o^{2}(1 - \frac{n \breve{\Phi} R_o^{\,n-1}}{(1+\breve{\Upsilon} R_o^{n})^{2}})}\right\}^{\Theta}-\frac{3W_{01}}{A_o^{2}B^{2}_o(1 - \frac{n \breve{\Phi} R_o^{\,n-1}}{(1+\breve{\Upsilon} R_o^{n})^{2}})}
\\\label{34a}&\times\left\{\frac{A'_o}{A_o}+\frac{2B'_o}{3B_o}+\frac{1}{3r} +\frac{C'_o}{3C_o}\right\}-\frac{3}{r^2}\left\{\frac{A^{\Theta}_o}{A_o}+\frac{2B^{\Theta}_o}{3B_o}
+\frac{C^{\Theta}_o}{3C_o}\right\}\frac{W_{02}}{A^{2}_oB_o^{2}(1 - \frac{n \breve{\Phi} R_o^{\,n-1}}{(1+\breve{\Upsilon} R_o^{n})^{2}})}
,
\\\nonumber&T^{1\zeta}_{;\zeta}=\frac{1}{B_o^{2}\left\{1 - \frac{n \breve{\Phi} R_o^{\,n-1}}{(1+\breve{\Upsilon} R_o^{n})^{2}}\right\}}\left[\left\{\frac{B'_o}{B_o}+\frac{A'_o}{A_o}-\frac{\left\{1 - \frac{n \breve{\Phi} R_o^{\,n-1}}{(1+\breve{\Upsilon} R_o^{n})^{2}}\right\}^{'}}{\left\{1 - \frac{n \breve{\Phi} R_o^{\,n-1}}{(1+\breve{\Upsilon} R_o^{n})^{2}}\right\}}-\frac{\left\{1 - \frac{n \breve{\Phi} R_o^{\,n-1}}{(1+\breve{\Upsilon} R_o^{n})^{2}}\right\}^{'}}{\left\{1 - \frac{n \breve{\Phi} R_o^{\,n-1}}{(1+\breve{\Upsilon} R_o^{n})^{2}}\right\}}+\frac{1}{r}+\frac{C'_o}{C_o}\right\} P_{xxo}\right.
\\\nonumber&
+\frac{1}{r}\left\{\frac{2B^{\Theta}_o}{B_o}+\frac{A^{\Theta}_o}{A_o}-\frac{f^{\Theta}_{i0R}}{\left\{1 - \frac{n \breve{\Phi} R_o^{\,n-1}}{(1+\breve{\Upsilon} R_o^{n})^{2}}\right\}}-\frac{f^{\Theta}_{i0R}}{\left\{1 - \frac{n \breve{\Phi} R_o^{\,n-1}}{(1+\breve{\Upsilon} R_o^{n})^{2}}\right\}}+\frac{C^{\Theta}_o}{C_o}\right\}P_{xyo}+\frac{1}{r}P^{\Theta}_{xyo}-\left\{\frac{1}{r}
+\frac{B'_o}{B_o}\right\}P_{yyo}-\frac{\acute{C_o}}{C_o}P_{zzo}+\frac{1}{2}
\\\nonumber
&\left.\left\{+\frac{\left\{R_o-\frac{\breve{\Phi} R_o^{n}}{1+\breve{\Upsilon} R_o^{n}}\right\}\left\{1 - \frac{n \breve{\Phi} R_o^{\,n-1}}{(1+\breve{\Upsilon} R_o^{n})^{2}}\right\}^{'}}{\left\{1 - \frac{n \breve{\Phi} R_o^{\,n-1}}{(1+\breve{\Upsilon} R_o^{n})^{2}}\right\}}\right.\right.\left.\left.+\left\{R_o-\frac{\breve{\Phi} R_o^{n}}{1+\breve{\Upsilon} R_o^{n}}\right\}-R'_o\left\{1 - \frac{n \breve{\Phi} R_o^{\,n-1}}{(1+\breve{\Upsilon} R_o^{n})^{2}}\right\}\right\}\right]-\left\{\frac{W_{01}}{A_o^{2}B^{2}_o\left\{1 - \frac{n \breve{\Phi} R_o^{\,n-1}}{(1+\breve{\Upsilon} R_o^{n})^{2}}\right\}}\right\}^{.}\\\nonumber
&+\left\{\frac{W_{11}}{B_o^{2}\left\{1 - \frac{n \breve{\Phi} R_o^{\,n-1}}{(1+\breve{\Upsilon} R_o^{n})^{2}}\right\}}\right\}^{'}+\left\{\frac{W_{12}}{B_o^{2}\left\{1 - \frac{n \breve{\Phi} R_o^{\,n-1}}{(1+\breve{\Upsilon} R_o^{n})^{2}}\right\}}\right\}^{\Theta}+\left\{\frac{B^{\Theta}_o}{B_o}
+\frac{C^{\Theta}_o}{4C_o}+\frac{A^{\Theta}}{4A_o}\right\}\times
\frac{4W_{12}}{B_o^{2}\left\{1 - \frac{n \breve{\Phi} R_o^{\,n-1}}{(1+\breve{\Upsilon} R_o^{n})^{2}}\right\}}\\\nonumber&-\frac{C'_oW_{33}}{B_o^{2}C_o\left\{1 - \frac{n \breve{\Phi} R_o^{\,n-1}}{(1+\breve{\Upsilon} R_o^{n})^{2}}\right\}}+\frac{A'_oW_{00}}{A_oB_o\left\{1 - \frac{n \breve{\Phi} R_o^{\,n-1}}{(1+\breve{\Upsilon} R_o^{n})^{2}}\right\}}
+\left\{\frac{B_o^{'}}{B_o}+\frac{C'_o}{3C_o}+\frac{A_o^{'}}{3A_o}+\frac{1}{3r}\right\}\times
\frac{3W_{11}}{B_o^{2}\left\{1 - \frac{n \breve{\Phi} R_o^{\,n-1}}{(1+\breve{\Upsilon} R_o^{n})^{2}}\right\}}
\\\label{35a}&-\left\{\frac{1}{r}+\frac{B'_o}{B_o}\right\}\frac{W_{22}}{B_o^{2}\left\{1 - \frac{n \breve{\Phi} R_o^{\,n-1}}{(1+\breve{\Upsilon} R_o^{n})^{2}}\right\}}+\frac{1}{B_o^{2}(1 - \frac{n \breve{\Phi} R_o^{\,n-1}}{(1+\breve{\Upsilon} R_o^{n})^{2}})}\left[\left\{\frac{A'_o}{A_o}\right\}\varrho_o+P'_{xxo}\right].
\end{align}

The static contribution to the Ricci scalar for considered geometry is obtained in the following form as:
\begin{align}\nonumber
&R_o=\frac{1}{B_o}\left[\frac{1}{B_o}\left\{\left\{\frac{2B_o}{B_o}-\frac{2A_o^{'}}{A_o}+\frac{2C_o^{'}}{C_o}\right\}\frac{1}{r}
+\frac{2C_o^{''}}{C_o}\right\}-\frac{2}{B_o}\left\{\left\{\frac{B_o^{'}}{B_o}\right\}^{2}-\frac{B_o^{''}}{B_o}\right\}\right.
+\left\{\frac{1}{A_o}\left\{A_o^{''}+\frac{A_o^{'}C'_o}{C_oA_o}\right\}\right\}\frac{4}{B_o}\\\label{37a}&\left. +\frac{1}{r}\left\{\frac{1}{B_o}\left\{\frac{2A^{\Theta\Theta}_o}{A_o}+\frac{2B^{\Theta\Theta}_o}{B_o}\right\}
+\frac{1}{B_o}\left\{\left\{2C^{\Theta\Theta}_o+\frac{2A^{\Theta}_oC_o^{\Theta}}{A_o}\right\}\frac{1}{C_o}
-2\left\{\frac{B^{\Theta}_o}{B_o}\right\}^{2}\right\}\right\}\right]
,
\end{align}

\subsection{Non Static Configuration:}

To extract the time dependent contributions from the original conservation equations, as Eqs.~\eqref{10a} to \eqref{12a}, we apply the adopted linear perturbation scheme and obtain the following relations as:
\begin{align}\nonumber
&T^{0\zeta}_{;\zeta}=\frac{1}{A_o^{2}\left\{1 - \frac{n \breve{\Phi} R_o^{\,n-1}}{(1+\breve{\Upsilon} R_o^{n})^{2}}\right\}}\left[\dot{\bar{\varrho}}+\left\{2\left\{\frac{c}{2C_o}+\frac{b}{B_o}\right\}\varrho_o+\frac{1}{2}\left\{e\left\{1 - \frac{n \breve{\Phi} R_o^{\,n-1}}{(1+\breve{\Upsilon} R_o^{n})^{2}}\right\}-
\frac{2bR_o\left\{1 - \frac{n \breve{\Phi} R_o^{\,n-1}}{(1+\breve{\Upsilon} R_o^{n})^{2}}\right\}}{A_oB_o}\right\}\right.\right.
\\\label{45a}&\left.\left.+\frac{c}{C_o}P_{zzo}+\frac{b}{B_o}P_{yyo}+\frac{b}{B_o}P_{xxo}+D_{1}\right\}\dot{Y}\right]
,
\\\nonumber&T^{1\zeta}_{;\zeta}=\frac{1}{B^{2}_o\left\{1 - \frac{n \breve{\Phi} R_o^{\,n-1}}{(1+\breve{\Upsilon} R_o^{n})^{2}}\right\}}\left[\bar{P'_{xx}}-\varrho_oY\left\{\frac{a}{A_o}\right\}^{'}
+P_{xxo}Y\left\{\frac{a}{A_o}\right\}^{'}+\frac{A'_o}{A_o}\bar{\varrho}+\bar{P_{xx}}\frac{A'_o}{A_o}
+YP_{xxo}\right.\\\nonumber&\times\left\{\frac{c}{C_o}\right\}^{'}-\left\{\frac{c}{C_o}\right\}^{'}YP_{zzo}+\bar{P_{xx}}\frac{C'_o}{C_o}
-\frac{C'_o}{C_o}\bar{P_{zz}}+YP_{xxo}\left\{\frac{b}{B_o}\right\}^{'}
-P_{yyo}Y\left\{\frac{b}{B_o}\right\}^{'}+\bar{P_{xx}}\frac{B_o^{'}}{B_o}\\\nonumber&-\frac{\bar{P^{\Theta}_{xy}}}{r}-\bar{P_{yy}}\frac{B_o^{'}}{B_o}
+\frac{\bar{P_{xx}}}{r}
-\frac{\bar{P_{yy}}}{r}+2\left\{\frac{B^{\Theta}_o}{rB_o}+\frac{A^{\Theta}_o}{2rA_o}+\frac{C^{\Theta}}{2rC_o}\right\}\bar{P_{xy}}
+\frac{Y}{r}\left\{\frac{c}{C_o}+2\frac{b}{B_o}+\frac{a}{A_o}\right\}^{\Theta}P_{xyo}\\\nonumber&+\frac{2Yb}{B_o}
\left\{\frac{A'_o}{A_o}\varrho_o+\frac{\left\{1 - \frac{n \breve{\Phi} R_o^{\,n-1}}{(1+\breve{\Upsilon} R_o^{n})^{2}}\right\}^{'}}{2\left\{1 - \frac{n \breve{\Phi} R_o^{\,n-1}}{(1+\breve{\Upsilon} R_o^{n})^{2}}\right\}}P_{xxo}\right\}
-\left\{\frac{P_{yyo}}{r}+\frac{C'_o}{C_o}P_{zzo}\right\}\times\frac{2Yb}{B_o}
\\\label{46a}&\left.-\frac{Yb}{B_o}\left\{\left\{R_o-\frac{\breve{\Phi} R_o^{n}}{1+\breve{\Upsilon} R_o^{n}}\right\}^{'}
-R'_o\left\{1 - \frac{n \breve{\Phi} R_o^{\,n-1}}{(1+\breve{\Upsilon} R_o^{n})^{2}}\right\}-\frac{(R_o-\frac{\breve{\Phi} R_o^{n}}{1+\breve{\Upsilon} R_o^{n}})\left\{1 - \frac{n \breve{\Phi} R_o^{\,n-1}}{(1+ R_o^{n}\breve{\Upsilon})^{2}}\right\}^{'}}{1 - \frac{n \breve{\Phi} R_o^{\,n-1}}{(1+ R_o^{n}\breve{\Upsilon})^{2}}}\right\}\right]+D_{2}=0
,
\end{align}

After implementing the linear perturbation scheme on the conservation equations, we extract the nonstatic (time-dependent) parts of our contemplated anisotropic highly dense matter composition, whereas the metric functions and matter variables are classified as static background values and minor perturbations as defined by the time-dependent function $Y(t)$. Subsequently, these equations are divided into static and non-static components, with the latter governing the physical development of matter distribution. Equations \eqref{45a}-\eqref{46a} represent the entire perturbed scheme, including corrections related to the considered $f(R)$ model through the coupling term. Equation~\eqref{45a} specifies the conservation of energy associated with the compact structures, that is explicitly connected to the static context of density $\varrho_0$ and anisotropic stresses by the curvature-dependent perturbations. The appearance of higher order curvature terms corresponding to $\breve{\Phi}$ and $\breve{\Upsilon}$ considerably affect the gravitational relationships. These equations are related to the combined parts of the perturbed equations of motion and directly incorporate $Y$ and $\dot{Y}$, establishing the fact that they represent the physical response of that composition of matter.
These relationships demonstrate that temporal fluctuations of perturbation functions significantly impact the shear stresses and off diagonal anisotropic parts, while $(a,b,c)$ along with thier radial and angular derivatives, implies the significant relevance of anisotropy in evaluating the stability of developing composition, and subsequentally they include the perturbed forms of the principal stresses $(\bar{P}_{xx}, \bar{P}_{yy}, \bar{P}_{zz})$.
Consequently, non-static composition is regulated by a refined interaction between anisotropic pressures, greater curvature corrections, and the time-dependent development specified by $Y(t)$, which collectively ascertain the system's stability or possibility for physical collapse. We identified new curvature components $D_{1}$ and $D_{2}$ corresponding to the considered gravity; and values for these terms are described in appendices. Meanwhile, non-static component of Eq. \eqref{13a} has subsequent form:
\begin{align}\nonumber
&Ye=\frac{\ddot{Y}}{A_o}\left\{\frac{2c}{A_oC_o}-\frac{2b}{A_oB_o}\right\}+\frac{2}{B_o}\left[\frac{A'_oYC'_o}{A_oC_o}\left\{\left\{\frac
{c'}{C'_o}+\frac{a'}{A_o}-\frac{a}{A_o}-\frac{c}{C_o}\right\}\right.
\frac{1}{B_o}\right\}-\frac{1}{B_o}\left\{\frac{YaA_o^{''}}{A_o}-\frac{Ya^{''}}{A_o}\right\}
+\frac{Y}{rB_o}\left\{\frac{c}{C_o}\right.\\\nonumber
&\left.-\frac{aB_o-bA_o}{A_oB_o}\right\}^{'}-\frac{1}{B_o}
\left\{\frac{YbB_o^{''}}{B_o^{2}}-\frac{Yb^{''}}{B_o}\right\}+\frac{Yc}{C_oB_o}\left\{\frac{C_o^{''}}{C_o}-\frac{c^{''}}{c}\right\}
-2\frac{B_o^{'}}{B^{2}_o}\left\{\frac{Yb'}{B_o}
-\frac{YbB'_o}{B_o^{2}}\right\}-\frac{c}
{B_oC_o}\left\{\frac{YC^{\Theta\Theta}_o}{C_o}-\frac{Yc^{\Theta\Theta}}
{c}\right\}\\\nonumber&-\frac{1}{B_o}\left\{\frac{YbB^{\Theta\Theta}_o}{B_o^{2}}-\frac{Yb^{\Theta\Theta}}{B_o}\right\}-\frac{A^{\Theta\Theta}
_o}{B_oA_o}\left\{\frac{Ya}{A_o}\right.\left.\left.
-\frac{Ya^{\Theta\Theta}}{A^{\Theta\Theta}_o}\right\}
+\frac{A_o^{\Theta}C_o^{\Theta}}{C_oA_o}
\left\{\left\{\frac{Ya^{\Theta}}{A^{\Theta}_o}-\frac{Ya}{A_o}+\frac{Yc^{\Theta}}{C^{\Theta}_o}
-\frac{Yc}{C_o}\right\}\frac{1}{B_o}\right\}
+2\frac{B_o^{\Theta}}{B^{2}_o}\left\{\frac{b}{B_o}\right\}^{\Theta}
\frac{Y}{r^{2}}\right]\\\label{48a}&-\frac{2YbR_o}{B_o}
,
\end{align}
{The purpose of the perturbation scheme is to examine whether a self-gravitating anisotropic configuration, initially in hydrostatic equilibrium, remains stable under small departures from equilibrium or evolves toward collapse. For this reason, all metric functions, matter variables and the Ricci scalar are written as the sum of their static background values and first order perturbations, after substituting these quantities into the modified field and conservation equations, we retain only linear terms in the perturbation parameter. This procedure separates the background equilibrium configuration from the dynamical part responsible for the onset of instability, thus the temporal dependence of the perturbation is governed by the linearized dynamical equation.
More precisely, the temporal perturbation equation as expressed in \eqref{48a} is now written in the form
\begin{align}\label{49a}
\frac{\partial}{\partial t} \left\{\frac{\partial Y}{\partial t}\right\} -sY=0, \qquad s>0,
\end{align}
In this scenario, the function whose value is described in Appendix, serves as an approach of simplifying calculations and representing particular characteristics of the matter distribution. Whereas general solution of Eq.\eqref{49a} is
\begin{align}\label{49aa}
Y(t)=\Upsilon_{1}e^{\sqrt{s}\,t}+\Upsilon_{2}e^{-\sqrt{s}\,t}.
\end{align}
Where $\Upsilon_1$ and $\Upsilon_2$ are are factors having dependence on preliminary constraints of the axially symmetric composition, while the constants $\Upsilon_1$ and $\Upsilon_2$ were influenced by the solution's (unstable) or steady increase over time. The choice of $\Upsilon_1$ and $\Upsilon_2$ can result oscillations within extremely dense axially symmetric system if $s < 0 $, a non-sustainable outcome in this compact configuration.
For a collapsing configuration which is assumed to be initially in equilibrium in the remote past, we impose
$$
Y(t)\rightarrow 0 \quad \text{as} \quad t\rightarrow -\infty .
$$
This condition removes the exponentially divergent branch \(e^{-\sqrt{s}t}\), and hence we set \(\Upsilon_{2}=0\). Choosing \(\Upsilon_{1}=-1\) fixes the sign convention for the collapsing perturbation and gives
\begin{align}\label{50a}
Y(t)=-e^{\sqrt{s}\,t}.
\end{align}
Thus, it is the physically selected collapsing branch of the general solution given above.}

\subsection{Collapse Equation}

To formulate the collapse equation, we relate the perturbations in the principal stresses to the perturbed energy density through the adiabatic index $\Gamma$. So, we establish the interaction between three dynamical parameters: energy density, pressures, and fluid stiffness, that is $\Gamma$ in subsequent form:
\begin{align}\label{51a}
\bar{P}_{k}=\frac{\bar{\varrho}P_{k0}}{\varrho_o+P_{k0}}\Gamma.
\end{align}
In this scenario, the previously mentioned connection may be derived with additional directions after inserting $k=xx,xy, yy,zz$.
Its broad implementation arises from its potential to represent the essential thermodynamic and physical characteristics of highly dense structures including supernova cores, neutron stars, and anisotropic compact matter configurations. The Harrison et al. EOS has been determined to sufficiently
dynamic and efficient for the evaluation of stability and perturbation analysis in various modified gravity frameworks, despite its being derived in GR. It incorporates a standard linear perturbation approach to evaluate the physical stability of fluid distribution, regardless of seeking a completely distinct EoS formalism.

In the context of $f(R)$ theory, additional structural parameters influence the gravitational interactions and results in non-conservation consequences, however, the dynamical evaluation of matter's internal thermodynamic behaviour remains effectively feasible through the consideration of EoS suggested by Harrison et al. The linearly correlated behaviour of pressure components to energy density perturbations based on $\Gamma$ provides basis for our considered perturbation methodology in the stability analysis within the framework of GR and revised gravity theories, however, the Harrison et al. EoS in the context of axially symmetric matter  distribution, implicitly incorporates anisotropy through the contemplation of $P_{xx},$ and $P_{yy},$ along with $P_{zz}$, and off-diagonal components $P_{xy}$.  Furthermore, the role of this EoS in the formalism of modified gravity theories is studied in various research articles, considering the significance of adiabatic index in determining the restraints for dynamical instability (e.g., \cite{sharif2015instability,yousaf2024unstable,ur2024dynamically}).
Integration of  Eq.\eqref{45a} results,
\begin{align}\nonumber
\bar{\varrho}&=-\frac{Y}{2}\left\{\left\{\frac{c}{2C_o}+\frac{b}{B_o}\right\}\varrho_o-
\frac{2bR_o\left\{1 - \frac{n \breve{\Phi} R_o^{\,n-1}}{(1+\breve{\Upsilon} R_o^{n})^{2}}\right\}}{A_oB_o}+\frac{2c}{C_o}P_{zzo}+e\left\{1 - \frac{n \breve{\Phi} R_o^{\,n-1}}{(1+\breve{\Upsilon} R_o^{n})^{2}}\right\}+\frac{2b}{B_o}P_{yyo}+\frac{2b}{B_o}P_{xxo}+2D_{1}\right\}
\end{align}
The conservation equations play a crucial role in establishing the stability constraints in N and pN domains. We now determine collapse equation in within the formalism of Hu-Sawicki framework after the insertion of  Eqs.~\eqref{51a}  Eqs.~\eqref{51a} into non-static components of conservation equation~\eqref{46a}, consequently we obtain the collapse equation as:
\begin{align}\nonumber&
\Gamma\left\{\frac{P_{xxo}}{\varrho_o+P_{xxo}}\right\}'\bar{\varrho}'-Y(P_{xxo}+\varrho_o)\left\{\frac{a}{A_o}\right\}'
+\frac{A_o^{'}}{A_o}\left\{\Gamma\frac{P_{xxo}}{\varrho_o+P_{xxo}}+1\right\}\bar{\varrho}\\\nonumber&
-Y(P_{zzo}-P_{zzo})\left\{\frac{c}{C_o}\right\}'
-\Gamma\frac{2C_o'}{C_o}\left\{\frac{P_{zzo}}{2\varrho_o+2P_{zzo}}-\frac{P_{xxo}}{2\varrho_o+2P_{xxo}}\right\}\bar{\varrho}-Y(P_{yyo}-P_{xxo})
\left\{\frac{b}{B_o}\right\}'
\\\nonumber&-2\Gamma\left\{\frac{P_{yyo}}{2\varrho_o+2P_{yyo}}-\frac{P_{xxo}}{2\varrho_o+2P_{xxo}}\right\}\left[\frac{B_o^{'}}{B_o}+\frac{1}{r}\right]
\bar{\varrho}+\frac{YP_{xyo}}{r}\left\{\frac{c}{C_o}+\frac{2b}{B_o}
+\frac{a}{A_o}\right\}^{\Theta}
\\\nonumber&+\Gamma\frac{1}{r}\left\{\frac{P_{xyo}}{\varrho_o+P_{xyo}}\right\}^{\Theta}[\bar{\varrho}]^{\Theta}
-\Gamma\frac{Y}{r}\left\{\frac{P_{xyo}}{\varrho_o+P_{xyo}}\right\}
\left\{\frac{A_o^{\Theta}}{2A_o}+\frac{B_o^{\Theta}}{B_o}+\frac{C_o^{\Theta}}{2C_o}\right\}\bar{\varrho}
+\frac{Yb}{B_o}\left\{\frac{2A'_o}{A_o}\varrho_o+P_{xxo}\right.
\\\nonumber&\times\left.\frac{\left\{1 - \frac{n \breve{\Phi} R_o^{\,n-1}}{(1+\breve{\Upsilon} R_o^{n})^{2}}\right\}^{'}}{1 - \frac{n \breve{\Phi} R_o^{\,n-1}}{(1+\breve{\Upsilon} R_o^{n})^{2}}}\right\}-\frac{2Yb}{B_o}\left\{\frac{1}{r}P_{yyo}+\frac{C'_o}{C_o}P_{zzo}\right\}-\frac{Yb(R_o-\frac{\breve{\Phi} R_o^{n}}{1+\breve{\Upsilon} R_o^{n}})}
{B_o}\left\{\frac{\left\{R_o-\frac{\breve{\Phi} R_o^{n}}{1+\breve{\Upsilon} R_o^{n}}\right\}^{'}}{R_o-\frac{\breve{\Phi} R_o^{n}}{1+\breve{\Upsilon} R_o^{n}}}-\frac{R'_o(1 - \frac{n \breve{\Phi} R_o^{\,n-1}}{(1+\breve{\Upsilon} R_o^{n})^{2}})}{R_o-\frac{\breve{\Phi} R_o^{n}}{1+\breve{\Upsilon} R_o^{n}}}\right.\\\label{57a}&\left.-\frac{\left\{1 - \frac{n \breve{\Phi} R_o^{\,n-1}}{(1+\breve{\Upsilon} R_o^{n})^{2}}\right\}^{'}}{1 - \frac{n \breve{\Phi} R_o^{\,n-1}}{(1+\breve{\Upsilon} R_o^{n})^{2}}}\right\}+D_{2}=0.
\end{align}
This process combines both perturbed energy density and pressure components with the conservation equation, resulting the specific form of conservation equation described in Eq.~\eqref{57a}, whereas, the derived result  determines the development of small perturbations and plays a crucial role in establishing the stability constraints for our considered axially symmetric compact structures. The interaction between these competing impacts establish the fact that matter composition maintains stability or collapses, specifically, the ratios $\frac{P_{ii0}}{\varrho_o+P_{ii0}}$ illustrate the importance of anisotropic stresses in determining the actual compressibility of the fluid composition. Furthermore, the greater order curvature terms corresponding to Hu-Sawicki model emerge from its particular factor along with its derivatives, that revise gravitating interactions and pressure gradients, so it changes the value of $\Gamma$ for determining stability , while the subsequent terms specify the remaining perturbative impacts resulting from modified gravity.

\section{Stability Domain:}

Now, we discuss collapse equation parameters in N and pN domains. Furthermore, we assess the stable or unstable features of compact objects in the formalism of modified  $f(R)$ model. The adiabatic index $\Gamma$ has significant relevance in interpreting the stable behaviour of axially symmetric compact objects as illustrated in earlier derived collapse equation.  In following subsections, we will discuss N and pN approaches relate to our considered axially symmetric compact structures.

\subsection{Newtonian Domain:}

Now, collapse equation will be derived in N domain after the assumption of specific values, such as $B_o=1=A_o~,~C_o=r$. Furthermore, we determine the formalism for $\Gamma$ employing the earlier established N collapse equation, that basically specifies the instability criterion after the consideration of particular constraints, in addition, we analyze the interaction between surface pressure and static part of energy density $(\varrho_o$) as $~P_{k0}<<\varrho_o$ ;$~k=~xx,~yy,~zz,~xy$.
Applying these constraints to earlier derived collapse equation in N domain results the following:
\begin{align}\nonumber
&-Y\varrho_oa'+\Gamma\frac{P_{zzo}}{r}\left\{\frac{2br+c}{r}\right\}Y+(P_{xxo}-P_{zzo})Y\left\{\frac{c}{r}\right\}'+\frac{b\left\{1 - \frac{n \breve{\Phi} R_o^{\,n-1}}{(1+\breve{\Upsilon} R_o^{n})^{2}}\right\}^{'}}{1 - \frac{n \breve{\Phi} R_o^{\,n-1}}{(1+\breve{\Upsilon} R_o^{n})^{2}}}YP_{xxo}
\\\nonumber&-\Gamma\frac{P_{xxo}}{r}
\left\{\frac{2br+c}{r}\right\}Y+b'(P_{xxo}-P_{yyo})Y-\Gamma\frac{1}{r}\left\{P_{xyo}\left\{\frac{2br+c}{r}\right\}\right\}^{\Theta}Y
\\\nonumber&+\frac{P_{xyo}}{r}\left\{\frac{2br+c+ar}{r}\right\}^{\Theta}Y-\frac{2b}{r}(P_{yyo}+P_{zzo})Y+b\left\{R_o-\frac{\breve{\Phi} R_o^{n}}{1+\breve{\Upsilon} R_o^{n}}\right\}\left\{\frac{R'_o(1 - \frac{n \breve{\Phi} R_o^{\,n-1}}{(1+\breve{\Upsilon} R_o^{n})^{2}})}{R_o-\frac{\breve{\Phi} R_o^{n}}{1+\breve{\Upsilon} R_o^{n}}}\right.
\\\label{58a}&\left.-\frac{\left\{R_o-\frac{\breve{\Phi} R_o^{n}}{1+\breve{\Upsilon} R_o^{n}}\right\}^{'}}{R_o-\frac{\breve{\Phi} R_o^{n}}{1+\breve{\Upsilon} R_o^{n}}}+\frac{\left\{1 - \frac{n \breve{\Phi} R_o^{\,n-1}}{(1+\breve{\Upsilon} R_o^{n})^{2}}\right\}^{'}}{1 - \frac{n \breve{\Phi} R_o^{\,n-1}}{(1+\breve{\Upsilon} R_o^{n})^{2}}}\right\}Y+D_{2N}=0,
\end{align}
The insertion of these interactions in collapse equation and contemplating the dominating factors, we study the impacts of curvature terms, anisotropic fluid dispersion, and metric potentials on stability of compact structure. The study yields the subsequent condition for adiabatic index $\Gamma$:
\begin{align}\label{59a}&
\Gamma^{N}<\frac{\Omega^{N}_1}{\Omega^{N}_2}.
\end{align}
in which, $\Omega^{N}_1=\left[\varrho_oa'-P_{1}\left\{\frac{c}{r}\right\}'-P_{2}b^{'}-\frac{P_{xyo}}{r}\left\{\frac{2br+c+ar}{r}\right\}^{\Theta}
-\frac{b\left\{R_o-\frac{\breve{\Phi} R_o^{n}}{1+\breve{\Upsilon} R_o^{n}}\right\}^{'}}{R_o-\frac{\breve{\Phi} R_o^{n}}{1+\breve{\Upsilon} R_o^{n}}}P_{xxo}-b(R_o-\frac{\breve{\Phi} R_o^{n}}{1+\breve{\Upsilon} R_o^{n}})F-D_{2N}\right]r$
and
$\Omega^{N}_2=\left\{\frac{2br+c}{r}\right\} (P_{zzo}- P_{xxo})-\left\{\left\{\frac{2br+c}{r}\right\} P_{xyo}\right\}^{\Theta}$.

{Expression~\eqref{59a} is obtained from the N limit of the collapse equation after substituting the perturbed form of the matter variables through the Harrison-Wheeler type relation for the considered axially symmetric anisotropic configuration. It gives the critical value of the adiabatic index \(\Gamma\), which measures the stiffness of the fluid against compression. In the compact notation used in the manuscript, the N instability condition may be written as
\[
\Gamma < \Gamma_{\rm crit}^{N},
\qquad
\Gamma_{\rm crit}^{N}=\frac{\Omega^{N}_{1}}{\Omega^{N}_{2}},
\]
where \(\Omega^{N}_{1}\) collects the effective destabilizing contributions, including the background density gradient, pressure anisotropy, shear stress and Hu-Sawicki curvature corrections which appear due to our considered $f(R)$ gravity, whereas \(\Omega^{N}_{2}\) contains the restoring pressure response terms generated by the perturbation of the anisotropic stresses. Therefore, expression~\eqref{59a} should not be read as an arbitrary algebraic inequality; rather, it represents the threshold separating stable and unstable configurations in the N approximation. The physical interpretation is the following. If the stiffness of the fluid, measured by \(\Gamma\), is smaller than the critical value \(\Gamma_{\rm crit}^{N}\), then the pressure response is not strong enough to counterbalance the combined attractive and curvature induced destabilizing effects, and the system becomes dynamically unstable. Conversely, for
\[
\Gamma > \Gamma_{\rm crit}^{N},
\]
the pressure reaction is sufficiently strong to resist collapse, as well as the configuration remains dynamically stable within the N approximation. The equality
\[
\Gamma = \Gamma_{\rm crit}^{N}
\]
corresponds to the marginally stable state. We have also clarified that this interpretation assumes the physically relevant case \(\Omega^{N}_{2}>0\). If \(\Omega^{N}_{2}<0\), the direction of the inequality must be reversed, and the corresponding model parameters must be examined separately for physical consistency or such cases require separate physical admissibility checks. In the general relativistic limit, obtained by switching off the Hu-Sawicki correction terms, the same structure reduces to the standard GR-type instability condition, showing that the present result is a modified gravity extension of the usual Chandrasekhar-type stability criterion.}

Several factors such as the anti-gravitational force, energy density, pressure anisotropy, additional curvature impacts  and intrinsic modifications related to $f(R)$ gravity formalism influence the instability criterion for axially symmetric compact structures. The system remains unstable as long as the derived inequality satisfy the constraints. The associated analysis offers a few crucial underlying insights into the equilibrium of various forces acting within the matter.
(a) Inequality ~\eqref{59a}, implies the dependence of stability on relationship between pressure gradiants, anti-gravitating factors along with the gravitational forces.
(b) Specifically, when
$
\left|\Omega^{N}_1\right|
$
matches
$
\left|\Omega^{N}_2\right|,
$
the celestial formation establishes hydrostatic equilibrium, which specifies an ideal balance between gravitational, inner pressure, and antigravitating forces.
(c)) If the gravitational factors in the numerator of inequality~\eqref{59a} dominate over
$
\left|\Omega^{N}_2\right|,
$
the matter composition retains its stability and prevents the phenomenon of gravitational collapse.
(d) In this context, the criteria for stability $\Gamma > 1.33$, is  fulfilled, which guarantees that the competing impacts of primary pressure components and the gravitating forces maintain the balanced position.
(e) In this formalism, both material functions and {effective curvature source terms} directly plays a role in establishing inequality defined by $\Gamma$, whereas significance of $\Gamma$ in determining the range of instability illustrates its importance as a diagnostic parameter for self-gravitational matter compositions.
The extent and persistency of instability are determined by the {effective curvature source terms}, resulting from additional curvature interactions in $f(R)$ theory, as well as the anisotropic pressure components and variable energy density. Moreover, deviations in fluid density and directional pressures cause diverse perturbative responses expressed through $\Gamma$ in distinct directions. As a result, the instability persists all through the development of matter, as all factors involved in the resulting inequality have a significant impact on physical behaviour of the restrained axially symmetric fluid distribution.

It is important to mention that the inequality for $\Gamma$ is determine by correction terms and material functions. This demonstrates the dependence of adiabatic index on basic structural features of matter, including anisotropic pressure components and energy density.  In this context, it can be asserted that the positivity of each numerator term corresponds to the emergence of instability. So, subsequent conditions should be satisfied: $P_{yyo}<P_{xxo}~,~P_{xxo}>P_{zzo}$ and and stability is endorsed by correction terms, that impair the consistency of compact systems and lead to unstable matter composition in N domain.
here, $P_{1},~P_{2}$ and $F$ are some short notations specified in subsequent form,
$$P_{1}=P_{xxo}-P_{zzo} ,\quad P_{2}=P_{xxo}-P_{yyo} ,\quad F=\left\{\frac{R'_o(1 - \frac{n \breve{\Phi} R_o^{\,n-1}}{(1+\breve{\Upsilon} R_o^{n})^{2}})}{R_o-\frac{\breve{\Phi} R_o^{n}}{1+\breve{\Upsilon} R_o^{n}}}-\frac{\left\{R_o-\frac{\breve{\Phi} R_o^{n}}{1+\breve{\Upsilon} R_o^{n}}\right\}^{'}}{R_o-\frac{\breve{\Phi} R_o^{n}}{1+\breve{\Upsilon} R_o^{n}}}+\frac{\left\{1 - \frac{n \breve{\Phi} R_o^{\,n-1}}{(1+\breve{\Upsilon} R_o^{n})^{2}}\right\}^{'}}{1 - \frac{n \breve{\Phi} R_o^{\,n-1}}{(1+\breve{\Upsilon} R_o^{n})^{2}}}\right\}.$$

\subsection{Post Newtonian Domain:}

Now, instability of the considered system is studied in pN domain. The constraints
such as $B_o=1-\frac{m_o}{r_{2}}, ~C_o=r,
~A_o=1+\frac{m_o}{r_{2}}$ are considered to derive the collapse equation in this regard. The terms of order one and avoiding higher order of the term $\frac{m_o}{r_{2}}$
are considered in collapse equation. This technique can be useful in understanding the primary causes for instability of contemplated system.
These factors are characterized as:
The considered formalism assists us to determine and evaluate the  primary variables contributing to the development of instability which is mainly driven by three interrelated factors: the first one is curvature modifications implemented by the revised gravitational formalism, the other is dynamical characteristics of anisotropic fluid, and third one is structural attributes specified by the metric potentials.
All of these components serves separately and collectively to the physical development of compact matter distribution, described as : (a) We contemplate a limited but dynamically significant class of axially symmetric, highly dense compositions that involves the extra curvature terms resulting from revised gravity framework impact the gravitational dynamics of self-gravitational structures.
(b) These greater-order curvature impacts not only affect the gravitational force but also the dynamics of compact matter,  implying a substantial contribution in establishing the stability of celestial structures. (c) The anisotropic matter variables, particularly the energy density and pressure components, serve as important parameters of the stability domain for self-gravitational matter compositions. (d) Any fluctuations in these parameters can disrupt the balanced position between gravitational forces and inner stresses, which could give rise to instability or gravitational collapse. (e) The metric potentials describe the fundamental structural configuration of spacetime and play an important role in determining the stability of the system. (f) The perturbed form of these metric functions implies direct impact on spatial curvature and gravitational potential, influencing the evolving behaviour of highly dense matter distribution. (g) After considering a systematic but restricted context, we identify the essential physical characteristics of the model by reducing extraneous complications caused by higher order curvature terms. (h) The determined collapse equation provides a comprehensive interpretation for the role of anisotropy, additional curvature terms, and structural configuration of spacetime in determining stability constraints for axially symmetric compact structures. We are intended to provide comprehensive description for the physical factors causing instability in non-static matter compositions, as well as improve the theoretical analysis of their behaviour within the context of alternative gravcity theories. In order to study the physical behaviour in more plausible domain, we evaluate the collapse equation in the framework of pN domain, however, after employing the above-mentioned hypothesis, the determined collapse equation takes the following form
\begin{align}\nonumber
&-\Gamma Y\left\{\frac{P_{xxo}}{\varrho_o+P_{xxo}}\right\}^{'}\left[\left\{2\left\{b-\frac{bm_o}{r_{2}}+\frac{c}{2r}\right\}\right\}\varrho_o+(P_{xxo}+P_{yyo})\left\{b
-\frac{bm_o}{r_{2}}\right\}+\frac{1}{2}e(1 - \frac{n \breve{\Phi} R_o^{\,n-1}}{(1+\breve{\Upsilon} R_o^{n})^{2}})\right.
\\\nonumber&\left.-bR_o(1 - \frac{n \breve{\Phi} R_o^{\,n-1}}{(1+\breve{\Upsilon} R_o^{n})^{2}})+\frac{c}{r}P_{zzo}+D_{1PN}\right]^{'}-Y(\varrho_o+P_{xxo})\left\{a+\frac{am_o}{r_{2}}\right\}'-Y\frac{m_o}{r_{2}^{2}}
\\\nonumber&\times\left[\left\{2\left\{b-\frac{bm_o}{r_{2}}+\frac{c}{2r}\right\}\right\}\varrho_o+(P_{xxo}+P_{yyo})\left\{b-\frac{bm_o}{r_{2}}\right\}
+\frac{1}{2}e(1 - \frac{n \breve{\Phi} R_o^{\,n-1}}{(1+\breve{\Upsilon} R_o^{n})^{2}})-bR_o(1 - \frac{n \breve{\Phi} R_o^{\,n-1}}{(1+\breve{\Upsilon} R_o^{n})^{2}})
\right.\\\nonumber&\left.+\frac{c}{r}P_{zzo}+D_{1PN}\right]+Y(P_{xxo}-P_{zzo})\left\{\frac{c}{r}\right\}'+Y(P_{xxo}-P_{yyo})\frac{bm_o}{r_{2}^{2}}
-\frac{m_oYP_{xxo}}{r_{2}^{2}(\varrho_o+P_{xxo})}\\\nonumber
&\times\Gamma \left[\left\{2\left\{b-\frac{bm_o}{r_{2}}+\frac{c}{2r}\right\}\right\}\varrho_o+(P_{xxo}+P_{yyo})\left\{b-\frac{bm_o}{r_{2}}\right\}
+\frac{1}{2}e(1 - \frac{n \breve{\Phi} R_o^{\,n-1}}{(1+\breve{\Upsilon} R_o^{n})^{2}})-bR_o(1 - \frac{n \breve{\Phi} R_o^{\,n-1}}{(1+\breve{\Upsilon} R_o^{n})^{2}})
\right.\\\nonumber&\left.+\frac{c}{r}P_{zzo}+D_{1PN}\right]+\Gamma\frac{Y}{r}\left\{\frac{P_{zzo}}{\varrho_o+P_{zzo}}-\frac{P_{xxo}}{\varrho_o+P_{xxo}}\right\}
\left[\left\{2\left\{b-\frac{bm_o}{r_{2}}+\frac{c}{2r}\right\}\right\}\varrho_o\right.
\\\nonumber&\left.+(P_{xxo}+P_{yyo})\left\{b-\frac{bm_o}{r_{2}}\right\}+\frac{1}{2}e(1 - \frac{n \breve{\Phi} R_o^{\,n-1}}{(1+\breve{\Upsilon} R_o^{n})^{2}})-bR_o(1 - \frac{n \breve{\Phi} R_o^{\,n-1}}{(1+\breve{\Upsilon} R_o^{n})^{2}})+\frac{c}{r}P_{zzo}+D_{1PN}\right]
\\\nonumber&+\Gamma Y\left\{\frac{P_{yyo}}{\varrho_o+P_{yyo}}-\frac{P_{xxo}}{\varrho_o+P_{xxo}}\right\}\left\{\frac{1}{r}-\frac{m_o}{r_{2}^{2}}\right\}
\left[\left\{2\left\{b-\frac{bm_o}{r_{2}}+\frac{c}{2r}\right\}\right\}\varrho_o\right.
\\\nonumber&\left.+(P_{xxo}+P_{yyo})\left\{b-\frac{bm_o}{r_{2}}\right\}+\frac{1}{2}e(1 - \frac{n \breve{\Phi} R_o^{\,n-1}}{(1+\breve{\Upsilon} R_o^{n})^{2}})-bR_o(1 - \frac{n \breve{\Phi} R_o^{\,n-1}}{(1+\breve{\Upsilon} R_o^{n})^{2}})+\frac{c}{r}P_{zzo}+D_{1PN}\right]\\\nonumber&
-\Gamma\frac{Y}{r}\left\{\frac{P_{xyo}}{\varrho_o+P_{xyo}}\right\}\left[\left\{2\left\{b-\frac{bm_o}{r_{2}}+\frac{c}{2r}\right\}\right\}\varrho_o
+(P_{xxo}+P_{yyo})\left\{b-\frac{bm_o}{r_{2}}\right\}\right.
\\\nonumber&\left.+\frac{1}{2}e\left\{1 - \frac{n \breve{\Phi} R_o^{\,n-1}}{(1+\breve{\Upsilon} R_o^{n})^{2}}\right\}-bR_o\left\{1 - \frac{n \breve{\Phi} R_o^{\,n-1}}{(1+\breve{\Upsilon} R_o^{n})^{2}}\right\}+\frac{c}{r}P_{zzo}+D_{1PN}\right]^{\Theta}+\left\{a-\frac{am_o}{r_{2}}
-\frac{2bm_o}{r_{2}}+\frac{c}{r}+2b\right\}^{\Theta}\\\nonumber&\times \frac{Y}{r}P_{xyo}+\frac{2Ybm_o}{r_{2}^{2}}\varrho_o+\frac{Yb\left\{1 - \frac{n \breve{\Phi} R_o^{\,n-1}}{(1+\breve{\Upsilon} R_o^{n})^{2}}\right\}^{'}}{1 - \frac{n \breve{\Phi} R_o^{\,n-1}}{(1+\breve{\Upsilon} R_o^{n})^{2}}}\left\{1-\frac{m_o}{r_{2}}\right\}P_{xxo}-\frac{2Yb}{r}(P_{yyo}-P_{zzo})
\left\{1-\frac{m_o}{r_{2}}\right\}
\\\label{60a}&-Yb\left\{R_o-\frac{\breve{\Phi} R_o^{n}}{1+\breve{\Upsilon} R_o^{n}}\right\}\left\{1-\frac{m_o}{r_{2}}\right\}\left\{\frac{\left\{R_o-\frac{\breve{\Phi} R_o^{n}}{1+\breve{\Upsilon} R_o^{n}}\right\}^{'}}{\left\{R_o-\frac{\breve{\Phi} R_o^{n}}{1+\breve{\Upsilon} R_o^{n}}\right\}}-\frac{R'_o(1 - \frac{n \breve{\Phi} R_o^{\,n-1}}{(1+\breve{\Upsilon} R_o^{n})^{2}})}{f_{i0}}
-\frac{\left\{1 - \frac{n \breve{\Phi} R_o^{\,n-1}}{(1+\breve{\Upsilon} R_o^{n})^{2}}\right\}^{'}}{1 - \frac{n \breve{\Phi} R_o^{\,n-1}}{(1+\breve{\Upsilon} R_o^{n})^{2}}}\right\}+D_{2PN}=0
\end{align}
In pN domain, terms are extended up to first order in $\frac{m_o}{r_{2}}$, considering weak gravitating fields and  gradual decrease in the motion of matter particles, afyter employing the earlier described hypothesis and including only pN order terms, the derived collapse equation is reduced to the form specified in Eq.~\eqref{60a}, that includes relativistic terms as well as higher order curvature impacts within the formalism of Hu-Sawicki model. In pN domain, the gravitational potential terms corresponding to $\frac{m_o}{r_{2}}$ imply direct impacts on pressure gradients and inner interactions, while the adiabatic index persists its significant role through the multiplied form of density perturbation terms and anisotropic stresses.
The combined consideration $\frac{P_{ii0}}{\varrho_o+P_{ii0}}$ implies the significant impacts on stability constraints caused by the anisotropic matter distribution, specifically by the differences $(P_{xxo}-P_{yyo})$ and $(P_{xxo}-P_{zzo})$, furthermore, the curvature contributions related to Hu-Sawicki model and corresponding radial derivatives are directly expressed in collapse equation that precisely modify the gravitational interactions and incorporate extra forces emerging from modified gravity formalism. The terms $D_{1PN}$ and $D_{2PN}$ contain greater order perturbative terms corresponding to the pN domain, assuring that all modifications beyond the specified approximation order are inevitably included, thus Eq.~\eqref{60a} describes the pN collapse constraint in general form within the formalism of Hu-Sawicki model. In order to maintain the axially symmetric composition over the pN context, extra curvature terms relating to modified gravity theory, materials, and metric functions are important. The above mentioned collapse assumption results an inequality that corresponds to the adiabatic index. The axially symmetric matter composition demonstrate instability, until following inequality is satisfied
\begin{align}\label{61a}&
\Gamma^{pN}<\frac{\Omega^{pN}_1}{\Omega^{pN}_2}.
\end{align}
where,
$\Omega^{pN}_1=\left\{\frac{c}{r}\right\}'P_{1}
+\eta_{(eff)}-N_{1}+\frac{N^{\Theta}_{2}}{r}P_{xyo}
+\frac{m_o}{r^{2}_{2}}\left\{2b\varrho_o+bP_{2}-K\right\}+D_{2PN}$, here $\eta_{(eff)}=-bM_{1}\left\{\frac{2P_{3}}{r}-\left\{\frac{\breve{\Phi} R_o^{n}}{1+\breve{\Upsilon} R_o^{n}}-R_o\right\}F-\left\{1 - \frac{n \breve{\Phi} R_o^{\,n-1}}{(1+\breve{\Upsilon} R_o^{n})^{2}}\right\}^{'}\left\{1 - \frac{n \breve{\Phi} R_o^{\,n-1}}{(1+\breve{\Upsilon} R_o^{n})^{2}}\right\}^{-1}P_{xxo}\right\}$
and
$\Omega^{pN}_2=K'K'_{1}+\frac{m_o}{r^{2}_{2}}K_{1}K+\frac{1}{r}\left\{K_{1}-K_{3}\right\}K+\frac{1}
{r}\left\{K_{1}-K_{2}\right\}\left\{1-\frac{m_o}{r_{2}}\right\}K+\frac{K^{\Theta}_{4}}{r}K$.
Every outcome in the preceding context fits into a system of various gravitational theories using the Hu-Sawicki methodology. We added various lengthy and repeating terms which are described in subsequent form in order to simplyfy our computations.
\begin{align}\nonumber
&K=\left\{2b-\frac{2bm_o}{r_{2}}+\frac{c}{r}\right\}\varrho_o+(P_{xxo}+P_{yyo})\left\{b-\frac{bm_o}{r_{2}}\right\}+\frac{1}{2}e\left\{1 - \frac{n \breve{\Phi} R_o^{\,n-1}}{(1+\breve{\Upsilon} R_o^{n})^{2}}\right\}-bR_o\left\{1 - \frac{n \breve{\Phi} R_o^{\,n-1}}{(1+\breve{\Upsilon} R_o^{n})^{2}}\right\}
\\\nonumber&+\frac{c}{r}P_{zzo}+D_{1PN},
\quad F=\left\{\frac{\left\{R_o-\frac{\breve{\Phi} R_o^{n}}{1+\breve{\Upsilon} R_o^{n}}\right\}^{'}}{\left\{R_o-\frac{\breve{\Phi} R_o^{n}}{1+\breve{\Upsilon} R_o^{n}}\right\}}-\frac{R'_o\left\{1 - \frac{n \breve{\Phi} R_o^{\,n-1}}{(1+\breve{\Upsilon} R_o^{n})^{2}}\right\}}{\left\{R_o-\frac{\breve{\Phi} R_o^{n}}{1+\breve{\Upsilon} R_o^{n}}\right\}}-\frac{\left\{1 - \frac{n \breve{\Phi} R_o^{\,n-1}}{(1+\breve{\Upsilon} R_o^{n})^{2}}\right\}^{'}}{\left\{1 - \frac{n \breve{\Phi} R_o^{\,n-1}}{(1+\breve{\Upsilon} R_o^{n})^{2}}\right\}}\right\},
\\\nonumber& P_{}=P_{xxo}-P_{zzo},\quad P_{2}=P_{xxo}-P_{yyo}, \quad P_{3}=P_{yyo}-P_{zzo}\\\nonumber& K_{1}=\frac{P_{xxo}}{\varrho_o+P_{xxo}},\quad K_{2}=\frac{P_{yyo}}{\varrho_o+P_{yyo}},\quad K_{4}=\frac{P_{xyo}}{\varrho_o+P_{xyo}},
\quad K_{3}=\frac{P_{zzo}}{\varrho_o+P_{zzo}},\\\nonumber& N_{1}=(\varrho_o+P_{xxo})\left\{a+\frac{am_o}{r_{2}}\right\}', \quad N_{2}=a-\frac{am_o}{r_{2}}+2b-\frac{2bm_o}{r_{2}}+\frac{c}{r},\\\nonumber&
M_{1}=1-\frac{m_o}{r_{2}},\quad M_{2}=1+\frac{m_o}{r_{2}}.
\end{align}

\subsection{GR-limits}

In the N regime, the restriction on the adiabatic index $\Gamma$ has already been obtained in compact form through Eqs.~\eqref{59a} and \eqref{61a}, while the quantities $\Omega^{N}_{1}$, $\Omega^{N}_{2}$, $\Omega^{pN}_{1}$, as well as $\Omega^{pN}_{2}$ collectively contain the contributions arising from the gravitational field, pressure anisotropy, and the modified curvature sector. In particular, Eqs.~\eqref{59a} and \eqref{61a} explicitly retain the dependence on the Hu-Sawicki parameters $\breve{\Phi}$ and $\breve{\Upsilon}$.
To recover the general relativistic sector from the full Hu-Sawicki $f(R)$ scenario, we impose the limiting procedure
$
\breve{\Phi}\rightarrow 0,
~
\breve{\Upsilon}\rightarrow 0.
$
This limit suppresses all non-Einsteinian curvature corrections generated by the Hu-Sawicki model and reduces the gravitational Lagrangian to the Einstein-Hilbert form, consequently, every term proportional to $\breve{\Phi}$ or $\breve{\Upsilon}$ disappears from the N and pN expressions for $\Gamma$. In the resulting reduced GR regime, the quantities $D^{N\text{-Red}}_{2}$ and $D^{pN\text{-Red}}_{2}$ contain only those contributions associated with the Einstein gravitational sector and the ordinary anisotropic matter distribution. Hence, the effective curvature source terms generated by the modified theory vanish, and the corresponding stability conditions reduce to their general relativistic forms:
\begin{equation}
f(R)\xrightarrow[\Phi,\Upsilon \to 0]{\ }\;R,
~~
D_{2NRed},D_{2pNRed}\xrightarrow[\Phi,\Upsilon \to0]{}\;0,
~~
\frac{\breve{\Phi} R_o^{n}}{1+\breve{\Upsilon} R_o^{n}}P_{xxo}\xrightarrow[\Phi,\Upsilon\to0]{}\;P_{xxo}.
\label{resp:eq:GRlimit}
\end{equation}

\subsubsection{N domain}
The results determined in GR can be obtained from inequality \eqref{59a} after considering the limit $\Phi,\Upsilon \rightarrow 0$. In his scenario, the equation relating to adiabatic index $\Gamma$ reduces to:
\begin{align}\label{59aa}&
\Gamma^{N}_{GR}<\frac{\Omega^{N}_{1GR}}{\Omega^{N}_{2GR}}.
\end{align}
where, $\Omega^{N}_{1GR}=\left[\varrho_oa'-P_{1}\left\{\frac{c}{r}\right\}'-P_{2}b^{'}-\frac{P_{xyo}}{r}\left\{\frac{2br+c+ar}{r}\right\}^{\Theta}
-\frac{bR_o^{'}}{R_o}P_{xxo}-bR_oF-D_{2NGR}\right]r$
and
$\Omega^{N}_{2GR}=\left\{\frac{2br+c}{r}\right\} (P_{zzo}- P_{xxo})-\left\{\left\{\frac{2br+c}{r}\right\} P_{xyo}\right\}^{\Theta}$

{As long as all terms in the inequality \eqref{59aa} satisfy conventional GR constraints, the instability related to the restrained non-static axially symmetric composition appears to be feasible.
Here, $D_{2NGR}$ represents the simplified form acquired in the limit $\Phi,\Upsilon \rightarrow 0$ within the context of GR. The instability \eqref{59aa} emphasize that the matter composition establishes hydrostatic equilibrium if the attractive force of gravity is effectively balanced by the outward pressure and the anti-gravitating effects. On the other hand, the dominance of gravitational pull corresponds to the stable matter configuration for which $\Gamma > 4/3$. In our manuscript, we evaluate the instability domain corresponding to the particular restrained axially symmetric matter distribution without applying the expansion-free condition. Our obtained findings reveal that the dynamical instability of compact systems is determined by the adiabatic index, in accordance with the predetermined theoretical assumptions \cite{chandrasekhar1964dynamical,herrera1989dynamical,chan1993dynamical,sharif2013stability}.

\subsubsection{pN domain}
Our study of specified non-static axially symmetric composition within the formalism of pN domain can be anticipated with conventional GR, through the assumption $\Phi,\Upsilon \rightarrow 0$ in instability \eqref{61a},  concludes the subsequent form:}

\begin{align}\label{61aa}&
\Gamma^{pN}_{GR}<\frac{\Omega^{pN}_{1GR}}{\Omega^{pN}_{2GR}}.
\end{align}
where,
$\Omega^{pN}_{1GR}=\left\{\frac{c}{r}\right\}'P_{1}
-bM_{1}\left\{\frac{2P_{3}}{r}-R_oF\right\}-N_{1}+\frac{N^{\Theta}_{2}}{r}P_{xyo}
+\frac{m_o}{r^{2}_{2}}\left\{2b\varrho_o+bP_{2}-K\right\}+D_{2PNGR}$,
and
$\Omega^{pN}_{2GR}=K'K'_{1}+\frac{m_o}{r^{2}_{2}}K_{1}K+\frac{1}{r}\left\{K_{1}-K_{3}\right\}K+\frac{1}
{r}\left\{K_{1}-K_{2}\right\}\left\{1-\frac{m_o}{r_{2}}\right\}K+\frac{K^{\Theta}_{4}}{r}K$.
Each result in the preceding context fits into a system of various gravitational theories using the Hu-Sawicki technique.
{The instability of the considered axially symmetric compact composition is retained as far as all the terms in above instability constraint are valid for the specified GR limitation. Eq.~\eqref{61aa} imposes certain limits on the adiabatic index $\Gamma$ to maintain dynamical equilibrium.
The limitation $\Phi,\Upsilon  \rightarrow 0$, $D_{2pNGR}$ implies the reduced form of terms in GR. We establish the instability constraint related to the specific axially symmetric matter distribution which reveal that the adiabatic index determines the emergence of instability, that is compatible with the standard stability restraint described in literature \cite{chandrasekhar1964dynamical,herrera1989dynamical,chan1993dynamical,sharif2013stability}.}
{Establishing the Chandrasekhar limit $\Gamma_{\rm crit}=4/3$ corresponding to isotropic fluid.
After obtaining the simplified form of results in GR (Eq.~\eqref{resp:eq:GRlimit}), the standard isotropic matter constraint can be applied through the consideration
\begin{equation}
P_{xxo}=P_{yyo}=P_{zzo}\equiv P_o,~P_{xyo}=0
~\Longrightarrow~
P_1=P_{zzo}-P_{xxo}=0,\;\; P_2=P_{xxo}-P_{yyo}=0,
\label{resp:eq:isotropic}
\end{equation}
that removes all anisotropic stresses and off-diagonal parts of pressure.   Within the context of these considerations, the instability constraints are simplified into standard GR results, which is consistent with Chandrasekhar's limitation for isotropic matter distribution in N and pN domains.}

\subsection{Dynamics of Axially Symmetric Collapsing Stars}

In last few years, interpretation of basic gravitational hypothesis, including cosmic censorship and hoop conjectures has prompted the analysis of non spherical collapse phenomenon, particularly, several studies have concluded that the collapsing fluid can develop into filament like structures, that can be precisely described through the cylindrical matter distributions. These compositions have significant relevance in analyzing the development of prestellar clouds, as well as the resulting creation of star \cite{borah2015theoretical}.
The evaluation of perturbative behaviour in compact structures has gained recognition as an important topic in astrophysics, provoking  considerable attention in theoretical and numerical contexts \cite{paramos2012gravitational,astashenok2015magnetic}, whereas stability evaluation of compact structures in GR offers major challenges, specifically when modelled within the context of idealistic consideration of spherical symmetry, however, distinct astrophysical configurations such as galaxy clusters, and DM halos, can be roughly described as spherically symmetric, physical cosmic configurations frequently deviate from complete symmetry.

In our case study, we investigate the physical development of self-gravitational, non-static, axially symmetric matter distributions using minimally connected tangential hyperbolic and linear formulations of  $f(R)$ revised gravity models. Our study emphasize on identifying the instability constraints of ansiotropic highly dense matter compositions, whereas the contemplated axially symmetric dispersion are expected to devoid of circular motion and thus, do not produce gravitational radiations \cite{herrera2015shearing}, whereas matter composition is defined by an anisotropic energy-momentum tensor, as stated in Eq.~\eqref{3a}, that leads to an interesting question:
\\
Can an isotropic matter dispersion produce gravitational waves?
\\
The earlier work on gravitational radiations, suggest that a relativistic dust cloud or a perfect matter composition  free from dissipative procedures do not emit gravitational waves as the fundamental EoS is essentially reversible, as the propagation of gravitational waves is an irreversible process, it demands entropy-producing procedures like heat dissipation, viscosity, so in the domain of astrophysics, the inclusion of a vorticity parameter is required for a self-gravitational composition to serve as a radiation system.

Our case study commences with a self-gravitational axially symmetric fluid distribution primarily in the state of hydrostatic equilibrium and then disrupted to evaluate the physical stability, as specified by Eqs.~\eqref{14a}, \eqref{23a} and \eqref{25a}, whereas the primary emphasize is on physical stability section, which investigate the behaviour of composition in the context of small perturbations about equilibrium, while thermal stability is neglected as the contemplated fluid distribution does not contain heat flux. After deriving the linear form of coservation and field equations with respect to radial perturbations, we determine constraints  establishing the emergence of instability, with the development of collapsing matter is affected by the interaction between stabilizing and destabilizing variables, suggesting that the matter composition can change its dynamics between stable and unstable states, based on its inner configuration.
The instability constraints in N and pN domains are crucial in interpreting the collapse behaviour of highly desne matter distribution, while gravitational collapse occurs in the result of interrupted hydrostatic equilibrium position of compact structures. The final fate of collapsing matter is determined by its initial mass, whereas in N physics, the stability of such compositions is defined by adiabatic index that corresponds to the ratio of energy density and pressure fluctuations \cite{chandrasekhar1984stars}. In the context of relativistic formalism, stability is also dependent to the compactness ratio, whereas in the case of axially symmetric compositions in revised gravity theory,  the conditions are more complicated because of the extra degrees of freedom emerging form interaction between matter and geometry in $f(R)$ gravity. In this manuscript, the consideration of $\Gamma$ as constant illustrate that aside from repeating GR-based findings, adiabatic index is additionally affected by the inputs of these interacting terms.

The classical stability analysis introduced by Chandrasekhar \cite{chandrasekhar1964dynamical} is extended here to a restricted class of anisotropic and axially symmetric self-gravitating configurations, within the adopted perturbative framework, the admissible stability ranges are determined by the conditions expressed in Eqs.~\eqref{59a}, \eqref{59aa}, \eqref{61a}, and \eqref{61aa}. These relations show that the critical value of $\Gamma$ is not universal, but depends on the background matter variables, directional stresses, metric perturbations, and higher curvature contributions. Consequently, hydrostatic stability in both the N and pN approximations is maintained only when $\Gamma$ satisfies the corresponding model dependent bounds. The present formulation neglects rotational, meridional, and electromagnetic effects, while previous studies indicate that magnetic fields may influence the generation and persistence of vorticity in compact objects as well as may thereby modify their dynamical evolution and gravitational radiation signatures. The simultaneous inclusion of magnetic stresses, rotation, and higher curvature corrections would therefore constitute a relevant extension of the present analysis, whereas our results provide a theoretical description of the response of high-density anisotropic matter to perturbations in a Hu-Sawicki inspired $f(R)$ gravitational background. In this work, the dynamical instability of a non-static, axially symmetric, anisotropic fluid configuration examined within the scenario of $f(R)$ gravity, whereas the gravitational sector is described by the Hu-Sawicki inspired function
$
f(R)
=
R
-
R_{c}
\frac{\Phi\left(R/R_{c}\right)^{n}}
{1+\Upsilon\left(R/R_{c}\right)^{n}},
$
where $\Phi$, $\Upsilon$, and $n$ regulate the nonlinear curvature corrections, while $R_{c}$ represents a characteristic curvature scale \cite{hu2007models,capozziello2008cosmography}. By defining
$
\breve{\Phi}
=
\Phi R_{c}^{\,1-n},
\qquad
\breve{\Upsilon}
=
\Upsilon R_{c}^{-n},
$
the model can equivalently be expressed as
$f(R)
=
R
-
\frac{\breve{\Phi}R^{n}}
{1+\breve{\Upsilon}R^{n}}.
$
This parametrization absorbs the explicit appearance of $R_{c}$ into the effective quantities $\breve{\Phi}$ and $\breve{\Upsilon}$ and is therefore more convenient for the perturbative treatment. The astrophysical corrections relevant to the present analysis are consequently governed by these effective parameter combinations. The principal outcomes of the analysis may be summarized as follows, first, the nonlinear curvature sector generates effective geometric source contributions that enter the field and conservation equations as well as modify the response of the fluid to perturbations. These terms influence the equilibrium conditions, the collapse equation, as well as the resulting instability bounds, second, the application of a linear perturbation scheme leads to explicit restrictions on $\Gamma$ in both the N and pN regimes, while the derived conditions show that the onset of instability is controlled by the joint action of ordinary gravitational attraction, pressure anisotropy, spatial gradients of the matter variables, and the curvature dependent source terms, whereas third, the axial anisotropy introduces additional stress components that are absent in an isotropic fluid and thereby changes the range of parameters compatible with equilibrium.

The higher-curvature terms should not be interpreted as universally repulsive or stabilizing, while their influence depends on their sign, magnitude, and coupling to the background matter and geometrical variables, however for certain parameter choices, they may counteract contraction and delay the departure from equilibrium, whereas in other regions they may strengthen the instability. The Hu-Sawicki sector therefore shifts the conventional critical bounds rather than producing a model independent enlargement of the stable domain. For comparison, gravitational collapse in GR is governed by the Einstein field equations together with the matter distribution, equation of state, and initial data, whereas depending on these ingredients, the evolution may lead to compact remnants, trapped regions, black hole formation, or singular behavior \cite{maslowski1968gravitational}. Modified gravity models extend this description by introducing additional degrees of freedom or curvature dependent terms that can appreciably alter the collapse process. Among these extensions, $f(R)$ gravity replaces the Ricci scalar in the Einstein--Hilbert action by a nonlinear function of $R$, while collapse analyses in such theories have shown that curvature corrections can modify both the instability thresholds as well as the possible final configurations \cite{cembranos2012gravitational,solanki2022gravitational,yousaf2025impacta}. In the Starobinsky model, the quadratic curvature term becomes important in high curvature environments and can substantially influence the effective pressure and gravitational response of the system \cite{astashenok2019gravitational}, whereas depending on the matter sector and boundary conditions, these corrections may soften the approach toward singular behavior or permit non-standard evolutionary phases. Further modifications arise in theories like $f(R,T)$ gravity, for example, the dependence on the trace of the energy momentum tensor generates additional effective forces whose influence on collapse depends strongly on the matter distribution \cite{abbas2019charged}. Related investigations in electromagnetic and $f(R,T,Q)$ scenarios demonstrate that curvature-matter couplings and charge can modify the equilibrium conditions and the rate of contraction \cite{sharif2023influence,sharif2023study}. Anisotropic compact configurations and stability of dense stellar models \cite{naseer2024extending}. Scalar tensor theories provide another mechanism through which gravitational collapse can depart from its general relativistic behavior \cite{matsuda1972gravitational,rudra2014gravitational,ziaie2024gravitational,duque2024emergent,farwa2025field}. Investigations of dissipative cylindrical systems further illustrate that heat flow, scalar dynamics, and anisotropy may collectively determine the collapse rate and instability conditions \cite{guha2014dissipative}, while in higher dimensional theories, such as Gauss--Bonnet and Lovelock gravity, quadratic and higher-order curvature invariants alter both the interior dynamics and the exterior spacetime of collapsing configurations \cite{rudra2011gravitational,hassannejad2023gravitational,bhatti2023novel,ditta2025exploring}. These contributions may delay trapped-surface formation, modify the singularity structure, or produce stability conditions that differ from their four-dimensional counterparts. Braneworld models similarly introduce bulk-induced corrections into the effective field equations on the brane and can therefore reshape the collapse dynamics \cite{wang2016black}. Taken together, these investigations show that the stability of compact objects is sensitive to pressure anisotropy, electric charge, scalar degrees of freedom, dimensional corrections, and nonlinear curvature effects. In the present model, anisotropic stresses and Hu-Sawicki curvature contributions act as additional dynamical ingredients that modify the critical conditions for departure from equilibrium. The derived N and pN inequalities therefore extend the classical Chandrasekhar picture to a more general axially symmetric environment. The results obtained here are consistent with the broader conclusion that the onset of gravitational instability is model dependent and cannot, in general, be characterized solely through the reference value $\Gamma=4/3$. These results demonstrate that anisotropy and higher-curvature corrections can appreciably modify the collapse dynamics and stability conditions of compact self-gravitating configurations.

\section{Concluding Remarks}

A general non-static axially symmetric spacetime may accommodate both rotational and meridional motions through the off-diagonal metric components \(dt\,d\Theta\) and \(dt\,d\varphi\). However, retaining these terms renders the corresponding gravitational equations highly nonlinear and considerably complicates the derivation of explicit instability conditions. For this reason, many investigations adopt restricted subclasses of axially symmetric geometries in which such mixed contributions are neglected. {In this work, we have investigated the dynamical instability of a restricted class of non-static, axially symmetric, self-gravitating fluid configurations in the scenario of a Hu-Sawicki inspired $f(R)$ gravity model. The matter distribution was described through an anisotropic energy-momentum tensor containing three unequal principal stresses together with an off-diagonal stress component, while by adopting a vorticity-free geometry, we formulated the corresponding conservation relations for the considered compact system. A linear perturbation scheme was applied to the metric functions, matter variables, and Ricci scalar about a static background configuration, whereas retaining terms up to first order in the perturbation parameter enabled us to separate the equilibrium and dynamical sectors as well as to derive the collapse equation governing the evolution of the perturbed fluid. The resulting expression incorporates contributions from the background energy density, directional pressures, anisotropic stresses, metric perturbations, as well as the higher curvature terms generated by the adopted $f(R)$ model.}

The instability conditions were subsequently obtained in terms of the adiabatic index $\Gamma$ under both N and pN approximations. These conditions demonstrate that the critical value of $\Gamma$ is not determined solely by the stiffness of the fluid, instead, it depends on the combined influence of the pressure anisotropy, spatial variation of the background variables, geometrical perturbations, and effective curvature source contributions. The N approximation provides the leading order instability range, whereas the pN treatment introduces additional relativistic corrections that further modify the admissible bounds on $\Gamma$. The role of the Hu-Sawicki curvature sector was examined by taking the limit in which its model parameters vanish, whereas in this limit, the modified gravitational function reduces to the Einstein-Hilbert form, and the corresponding instability conditions recover their general relativistic expressions. Moreover, after imposing
$P_{xx0}=P_{yy0}=P_{zz0}\equiv P_{0},~P_{xy0}=0,$ the anisotropic contributions disappear and the classical Chandrasekhar reference value is recovered for an effectively isotropic configuration. This limiting behavior provides a useful consistency check on the derived results. Overall, the present analysis demonstrates that the higher-curvature contributions arising in $f(R)$ gravity provide a more comprehensive description of the stability properties of compact stellar configurations. The derived results are consistent with earlier investigations \cite{herrera2012dynamical,sharif2015instability} and further clarify the role of modified gravitational effects in governing the onset of dynamical instability in anisotropic self-gravitating systems.

{Our analysis therefore shows that nonlinear curvature effects as well as anisotropic stresses can shift the conventional instability threshold of axially symmetric compact configurations. Depending on the signs and relative magnitudes of the effective source terms, these contributions may either oppose or promote the departure from equilibrium, hence, the onset of collapse in the present model is controlled by an interplay between the material properties of the fluid and the modified gravitational sector rather than by a universal value of the adiabatic index alone. The conclusions obtained here apply to the restricted configuration adopted in this study, for which meridional motion, rotation, dissipative transport, electromagnetic fields, and higher-order perturbations neglected. Incorporating these effects, together with a numerical examination of representative stellar profiles and viable ranges of the Hu-Sawicki parameters, would provide a natural extension of the present analysis.}


\section{Appendix}

{For notational convenience and to improve the readability of the derived equations, several lengthy combinations have been represented by compact auxiliary quantities, while their complete analytical forms are provided in the Appendix as}
\begin{align}\nonumber&
s=-\frac{R_ob}{B_o}\frac{A^2_o}{(\frac{b}{B_o}-\frac{c}{C_o})}-A^{2}_o\left[\frac{e}{2}-\frac{A'_o}{A_o}\left\{\frac{C'_o}{C_o}
\left\{\frac{c'}{B^2_oC'_o}-\frac{a}{A_oB^2_o}+\frac{a'}{B^2_oA'_o}-\frac{c}{C_oB^2_o}\right\}
-\left\{\frac{b'}{b}-\frac{B'_o}{B_o}\right\}\frac{bB_o^{'}}{B^{4}_o}+\left\{\frac{B_oc+C_ob}{B_oC_o}
\right.\right.\right.\\\nonumber
&\left.-\frac{a}{A_o}
\right\}^{'}\frac{1}{2B^2_or}+\frac{C_o^{''}}{C_o}\left\{\left\{\frac{c}{C_o}-\frac{c^{''}}{C_o^{''}}\right\}\frac{1}{B^2_o}\right\}
+\frac{B_o^{\Theta}}{r^{2}B_o^2}\left\{\frac{b}{B_o}\right\}^{\Theta}-\frac{C_o^{\Theta\Theta}}{B_0C_o}
\left\{\frac{c}{2C_o}
-\frac{c^{\Theta\Theta}}{2C^{\Theta\Theta}_o}\right\}
-\frac{A_o^{''}}{A_o}\left\{\left\{\frac{a}{2A_o}-\frac{a^{''}}{2A^{''}_o}\right\}\frac{1}{B_o}\right\}
\\\nonumber
&-\frac{B^{''}_o}{B_o}
\left\{\left\{\frac{b}{2B_o}-\frac{b^{''}}{2B^{''}_o}\right\}\frac{1}{B_o}\right\}-\frac{1}{B_o}\left\{\frac{B^{\Theta\Theta}_o}{B_o}\left\{\frac{b}{2B_o}-\frac{b^{\Theta\Theta}}{2B^{\Theta\Theta}_o}
\right\}-\frac{A^{\Theta\Theta}_o}{A_o}\left\{\frac{a}{2A_o}-\frac{a^{\Theta\Theta}}{2A^{\Theta\Theta}_o}\right\}\right\}
+\frac{A_o^{\Theta}}{A_o}\left\{\frac{C_o^{\Theta}}{C_o}\left\{\frac{c^{\Theta}}{2C^{\Theta}_o}+\frac{a^{\Theta}}{2A^{\Theta}_o}\right.\right.
\\\nonumber&\left.\left.\left.\left.-\frac{a}{2A_o}-\frac{c}{2C_o}\right\}\frac{1}{B_o}\right\}\right\}\frac{1}{(\frac{b}{B_o}
-\frac{c}{C_o})}\right].
\end{align}
{The quantities $D_i^{\prime s}$ denote the lengthy contributions that emerge in the nonstatic conservation equations, while in the N regime, the curvature corrections associated with the adopted modified gravity model introduce an additional term, represented by $D_{2N}$. This quantity acts as an effective dark source contribution to the N background and incorporates the influence of the higher order curvature sector on the dynamical evolution of the system, whereas its complete expression is given as follows:}
\begin{align}\nonumber&
D_{2N}=\frac{Y}{\left\{1 - \frac{n \breve{\Phi} R_o^{\,n-1}}{(1+\breve{\Upsilon} R_o^{n})^{2}}\right\}}\left[\left\{\frac{2-Ye n\breve{\Phi} R_o^{n-2}\left[\frac{(n-1)(1+\breve{\Upsilon} R_o^{n})-2n \breve{\Upsilon} R_o^{n}}{(1+\breve{\Upsilon} R_o^{n})^{3}}\right]\left\{1 - \frac{n \breve{\Phi} R_o^{\,n-1}}{(1+\breve{\Upsilon} R_o^{n})^{2}}\right\}^{'}}{\left\{1 - \frac{n \breve{\Phi} R_o^{\,n-1}}{(1+\breve{\Upsilon} R_o^{n})^{2}}\right\}^{2}}+\frac{Ye n\breve{\Phi} R_o^{n-2}\left[\frac{(n-1)(1+\breve{\Upsilon} R_o^{n})-2n \breve{\Upsilon} R_o^{n}}{(1+\breve{\Upsilon} R_o^{n})^{3}}\right]}{\left\{1 - \frac{n \breve{\Phi} R_o^{\,n-1}}{(1+\breve{\Upsilon} R_o^{n})^{2}}\right\}}\right.\right.\\\nonumber&\left.\left.-\frac{4Yb}{r}+\frac{Ye n\breve{\Phi} R_o^{n-2}\left[\frac{(n-1)(1+\breve{\Upsilon} R_o^{n})-2n \breve{\Upsilon} R_o^{n}}{(1+\breve{\Upsilon} R_o^{n})^{3}}\right]}
{r\left\{1 - \frac{n \breve{\Phi} R_o^{\,n-1}}{(1+\breve{\Upsilon} R_o^{n})^{2}}\right\}}\right\}P_{xxo}-\left\{2Yb+\frac{-Ye n\breve{\Phi} R_o^{n-2}\left[\frac{(n-1)(1+\breve{\Upsilon} R_o^{n})-2n \breve{\Upsilon} R_o^{n}}{(1+\breve{\Upsilon} R_o^{n})^{3}}\right]}{\left\{1 - \frac{n \breve{\Phi} R_o^{\,n-1}}{(1+\breve{\Upsilon} R_o^{n})^{2}}\right\}}\right\} P'_{xxo}\right.\\\nonumber
&+\left\{\frac{2-Ye n\breve{\Phi} R_o^{n-2}\left[\frac{(n-1)(1+\breve{\Upsilon} R_o^{n})-2n \breve{\Upsilon} R_o^{n}}{(1+\breve{\Upsilon} R_o^{n})^{3}}\right]\left\{1 - \frac{n \breve{\Phi} R_o^{\,n-1}}{(1+\breve{\Upsilon} R_o^{n})^{2}}\right\}^{'}}{\left\{1 - \frac{n \breve{\Phi} R_o^{\,n-1}}{(1+\breve{\Upsilon} R_o^{n})^{2}}\right\}^{2}}+\frac{\left\{Ye n\breve{\Phi} R_o^{n-2}\left[\frac{(n-1)(1+\breve{\Upsilon} R_o^{n})-2n \breve{\Upsilon} R_o^{n}}{(1+\breve{\Upsilon} R_o^{n})^{3}}\right]\right\}^{\Theta}}{\left\{1 - \frac{n \breve{\Phi} R_o^{\,n-1}}{(1+\breve{\Upsilon} R_o^{n})^{2}}\right\}}\right.\\\nonumber&\left.-\frac{Yb\left\{1 - \frac{n \breve{\Phi} R_o^{\,n-1}}{(1+\breve{\Upsilon} R_o^{n})^{2}}\right\}^{\Theta}}{\left\{1 - \frac{n \breve{\Phi} R_o^{\,n-1}}{(1+\breve{\Upsilon} R_o^{n})^{2}}\right\}}\right\}P_{xyo}
-\frac{\left\{1 - \frac{n \breve{\Phi} R_o^{\,n-1}}{(1+\breve{\Upsilon} R_o^{n})^{2}}\right\}^{\Theta}}{r\left\{1 - \frac{n \breve{\Phi} R_o^{\,n-1}}{(1+\breve{\Upsilon} R_o^{n})^{2}}\right\}}\bar{P}_{xy}-\frac{1}{r}\left\{2Yb+\frac{-Ye n\breve{\Phi} R_o^{n-2}\left[\frac{(n-1)(1+\breve{\Upsilon} R_o^{n})-2n \breve{\Upsilon} R_o^{n}}{(1+\breve{\Upsilon} R_o^{n})^{3}}\right]}{\left\{1 - \frac{n \breve{\Phi} R_o^{\,n-1}}{(1+\breve{\Upsilon} R_o^{n})^{2}}\right\}}\right\} P^{\Theta}_{xxo}\\\nonumber&-\frac{-Ye n\breve{\Phi} R_o^{n-2}\left[\frac{(n-1)(1+\breve{\Upsilon} R_o^{n})-2n \breve{\Upsilon} R_o^{n}}{(1+\breve{\Upsilon} R_o^{n})^{3}}\right]}{r\left\{1 - \frac{n \breve{\Phi} R_o^{\,n-1}}{(1+\breve{\Upsilon} R_o^{n})^{2}}\right\}}P_{yyo}+\frac{-Ye n\breve{\Phi} R_o^{n-2}\left[\frac{(n-1)(1+\breve{\Upsilon} R_o^{n})-2n \breve{\Upsilon} R_o^{n}}{(1+\breve{\Upsilon} R_o^{n})^{3}}\right]}{\left\{1 - \frac{n \breve{\Phi} R_o^{\,n-1}}{(1+\breve{\Upsilon} R_o^{n})^{2}}\right\}}P_{zzo}
\\\nonumber&
-\frac{\left\{1 - \frac{n \breve{\Phi} R_o^{\,n-1}}{(1+\breve{\Upsilon} R_o^{n})^{2}}\right\}^{'}}{\left\{1 - \frac{n \breve{\Phi} R_o^{\,n-1}}{(1+\breve{\Upsilon} R_o^{n})^{2}}\right\}}\bar{P}_{xx}+\frac{1}{2}\left\{\left\{Ye\left[1-\frac{n \breve{\Phi} R_o^{\,n-1}}{(1+\breve{\Upsilon} R_o^{n})^{2}}\right]\right\}^{'}+\frac{\left\{Ye\left[1-\frac{n \breve{\Phi} R_o^{\,n-1}}{(1+\breve{\Upsilon} R_o^{n})^{2}}\right]\right\}^{'}\left\{1 - \frac{n \breve{\Phi} R_o^{\,n-1}}{(1+\breve{\Upsilon} R_o^{n})^{2}}\right\}^{'}}{\left\{1 - \frac{n \breve{\Phi} R_o^{\,n-1}}{(1+\breve{\Upsilon} R_o^{n})^{2}}\right\}}\right.\\\nonumber&\left.-\frac{Ye n\breve{\Phi} R_o^{n-2}\left[\frac{(n-1)(1+\breve{\Upsilon} R_o^{n})-2n \breve{\Upsilon} R_o^{n}}{(1+\breve{\Upsilon} R_o^{n})^{3}}\right]\left\{R_o-\frac{\breve{\Phi} R_o^{n}}{1+\breve{\Upsilon} R_o^{n}}\right\}}{\left\{1 - \frac{n \breve{\Phi} R_o^{\,n-1}}{(1+\breve{\Upsilon} R_o^{n})^{2}}\right\}}-\frac{2-Ye n\breve{\Phi} R_o^{n-2}\left[\frac{(n-1)(1+\breve{\Upsilon} R_o^{n})-2n \breve{\Upsilon} R_o^{n}}{(1+\breve{\Upsilon} R_o^{n})^{3}}\right]}{\left\{1 - \frac{n \breve{\Phi} R_o^{\,n-1}}{(1+\breve{\Upsilon} R_o^{n})^{2}}\right\}^{2}}\times\left\{R_o-\frac{\breve{\Phi} R_o^{n}}{1+\breve{\Upsilon} R_o^{n}}\right\}\right.\\\nonumber&\left.\times
\left\{1 - \frac{n \breve{\Phi} R_o^{\,n-1}}{(1+\breve{\Upsilon} R_o^{n})^{2}}\right\}^{'}-2R'_o-Ye n\breve{\Phi} R_o^{n-2}\left[\frac{(n-1)(1+\breve{\Upsilon} R_o^{n})-2n \breve{\Upsilon} R_o^{n}}{(1+\breve{\Upsilon} R_o^{n})^{3}}\right]-Ye'\left\{1 - \frac{n \breve{\Phi} R_o^{\,n-1}}{(1+\breve{\Upsilon} R_o^{n})^{2}}\right\}\right.\\\nonumber&\left.\left.-\frac{-Ye n\breve{\Phi} R_o^{n-2}\left[\frac{(n-1)(1+\breve{\Upsilon} R_o^{n})-2n \breve{\Upsilon} R_o^{n}}{(1+\breve{\Upsilon} R_o^{n})^{3}}\right]\left\{R_o-\frac{\breve{\Phi} R_o^{n}}{1+\breve{\Upsilon} R_o^{n}}\right\}^{'}}{\left\{1 - \frac{n \breve{\Phi} R_o^{\,n-1}}{(1+\breve{\Upsilon} R_o^{n})^{2}}\right\}}\right\}\right]+\left\{2Yb+\frac{-Ye n\breve{\Phi} R_o^{n-2}\left[\frac{(n-1)(1+\breve{\Upsilon} R_o^{n})-2n \breve{\Upsilon} R_o^{n}}{(1+\breve{\Upsilon} R_o^{n})^{3}}\right]}{\left\{1 - \frac{n \breve{\Phi} R_o^{\,n-1}}{(1+\breve{\Upsilon} R_o^{n})^{2}}\right\}}\right\}
\\\nonumber&\times\left\{\frac{W_{01}}{\left\{1 - \frac{n \breve{\Phi} R_o^{\,n-1}}{(1+\breve{\Upsilon} R_o^{n})^{2}}\right\}}\right\}^{.}+\left\{\left\{2Yb+2Ya+\frac{-Ye n\breve{\Phi} R_o^{n-2}\left[\frac{(n-1)(1+\breve{\Upsilon} R_o^{n})-2n \breve{\Upsilon} R_o^{n}}{(1+\breve{\Upsilon} R_o^{n})^{3}}\right]}{\left\{1 - \frac{n \breve{\Phi} R_o^{\,n-1}}{(1+\breve{\Upsilon} R_o^{n})^{2}}\right\}}\right\}\frac{W_{01}}{\left\{1 - \frac{n \breve{\Phi} R_o^{\,n-1}}{(1+\breve{\Upsilon} R_o^{n})^{2}}\right\}}\right\}^{.}\\\nonumber&
-\left\{\frac{W_{11}}{\left\{1 - \frac{n \breve{\Phi} R_o^{\,n-1}}{(1+\breve{\Upsilon} R_o^{n})^{2}}\right\}}\right\}'\times\left\{2Yb+\frac{-Ye n\breve{\Phi} R_o^{n-2}\left[\frac{(n-1)(1+\breve{\Upsilon} R_o^{n})-2n \breve{\Upsilon} R_o^{n}}{(1+\breve{\Upsilon} R_o^{n})^{3}}\right]}{\left\{1 - \frac{n \breve{\Phi} R_o^{\,n-1}}{(1+\breve{\Upsilon} R_o^{n})^{2}}\right\}}\right\}-\left\{\left\{2Yb\right.\right.\\\nonumber&\left.\left.-\frac{Ye n\breve{\Phi} R_o^{n-2}\left[\frac{(n-1)(1+\breve{\Upsilon} R_o^{n})-2n \breve{\Upsilon} R_o^{n}}{(1+\breve{\Upsilon} R_o^{n})^{3}}\right]}{\left\{1 - \frac{n \breve{\Phi} R_o^{\,n-1}}{(1+\breve{\Upsilon} R_o^{n})^{2}}\right\}}\right\}
\frac{W_{11}}{\left\{1 - \frac{n \breve{\Phi} R_o^{\,n-1}}{(1+\breve{\Upsilon} R_o^{n})^{2}}\right\}}\right\}'-\left\{\left\{2Eb+\frac{-Ye n\breve{\Phi} R_o^{n-2}\left[\frac{(n-1)(1+\breve{\Upsilon} R_o^{n})-2n \breve{\Upsilon} R_o^{n}}{(1+\breve{\Upsilon} R_o^{n})^{3}}\right]}{\left\{1 - \frac{n \breve{\Phi} R_o^{\,n-1}}{(1+\breve{\Upsilon} R_o^{n})^{2}}\right\}}\right\}\right.\\\nonumber&\left.\times\frac{W_{12}}{\left\{1 - \frac{n \breve{\Phi} R_o^{\,n-1}}{(1+\breve{\Upsilon} R_o^{n})^{2}}\right\}}\right\}^{\Theta}
-\left\{\frac{W_{12}}{\left\{1 - \frac{n \breve{\Phi} R_o^{\,n-1}}{(1+\breve{\Upsilon} R_o^{n})^{2}}\right\}}\right\}^{\Theta}\left\{2Yb+\frac{-Ye n\breve{\Phi} R_o^{n-2}\left[\frac{(n-1)(1+\breve{\Upsilon} R_o^{n})-2n \breve{\Upsilon} R_o^{n}}{(1+\breve{\Upsilon} R_o^{n})^{3}}\right]}{\left\{1 - \frac{n \breve{\Phi} R_o^{\,n-1}}{(1+\breve{\Upsilon} R_o^{n})^{2}}\right\}}\right\}
-\left\{4\dot{Y}b+\dot{Y}a\right.\\\nonumber&\left.+\frac{c}{r}
\dot{Y}\right\}\times\frac{W_{01}}{\left\{1 - \frac{n \breve{\Phi} R_o^{\,n-1}}{(1+\breve{\Upsilon} R_o^{n})^{2}}\right\}}+\left\{3Yb'+Ya'+\frac{Yc'}{r}-\frac{Yc}{r^{2}}-\frac{8Yb}{r}
-\frac{2-Ye n\breve{\Phi} R_o^{n-2}\left[\frac{(n-1)(1+\breve{\Upsilon} R_o^{n})-2n \breve{\Upsilon} R_o^{n}}{(1+\breve{\Upsilon} R_o^{n})^{3}}\right]}{r\left\{1 - \frac{n \breve{\Phi} R_o^{\,n-1}}{(1+\breve{\Upsilon} R_o^{n})^{2}}\right\}}\right\}\\\nonumber&\times\frac{W_{11}}{\left\{1 - \frac{n \breve{\Phi} R_o^{\,n-1}}{(1+\breve{\Upsilon} R_o^{n})^{2}}\right\}}+\left\{4Yb^{\Theta}+Ya^{\Theta}+\frac{Yc^{\Theta}}{r}\right\}\times\frac{W_{12}}{\left\{1 - \frac{n \breve{\Phi} R_o^{\,n-1}}{(1+\breve{\Upsilon} R_o^{n})^{2}}\right\}}
-\frac{W_{22}}{r\left\{1 - \frac{n \breve{\Phi} R_o^{\,n-1}}{(1+\breve{\Upsilon} R_o^{n})^{2}}\right\}}\left\{rYb'-4Yb\right.\\\nonumber&\left.+\frac{Ye n\breve{\Phi} R_o^{n-2}\left[\frac{(n-1)(1+\breve{\Upsilon} R_o^{n})-2n \breve{\Upsilon} R_o^{n}}{(1+\breve{\Upsilon} R_o^{n})^{3}}\right]}{\left\{1 - \frac{n \breve{\Phi} R_o^{\,n-1}}{(1+\breve{\Upsilon} R_o^{n})^{2}}\right\}}\right\}
-\left\{\frac{c'}{2r}-\frac{2b}{r}-\frac{c}{2r}+\frac{e n\breve{\Phi} R_o^{n-2}\left[\frac{(n-1)(1+\breve{\Upsilon} R_o^{n})-2n \breve{\Upsilon} R_o^{n}}{(1+\breve{\Upsilon} R_o^{n})^{3}}\right]}{\left\{1 - \frac{n \breve{\Phi} R_o^{\,n-1}}{(1+\breve{\Upsilon} R_o^{n})^{2}}\right\}}\right\}\frac{YW_{33}}{1 - \frac{n \breve{\Phi} R_o^{\,n-1}}{(1+\breve{\Upsilon} R_o^{n})^{2}}}.
\end{align}
{The energy density distribution associated with the nonstatic sector in the N era can be written as:}
\begin{align}
\nonumber&\bar{\varrho}_{N}=-\frac{Y}{2}\left\{\left\{\frac{c}{2r}+b\right\}\varrho_o-2bR_o\left\{1 - \frac{n \breve{\Phi} R_o^{\,n-1}}{(1+\breve{\Upsilon} R_o^{n})^{2}}\right\}+\frac{2c}{r}P_{zzo}+e\left\{1 - \frac{n \breve{\Phi} R_o^{\,n-1}}{(1+\breve{\Upsilon} R_o^{n})^{2}}\right\}+2bP_{yyo}+2bP_{xxo}+2D_{1}\right\}
\end{align}
{This relation characterizes the energy density perturbation of the nonstatic sector in the N era, while it contains contributions from the background energy density $\varrho_o$, the principal pressure components $P_{xxo}$, $P_{yyo}$, and $P_{zzo}$, the background Ricci scalar $R_o$, as well as the additional quantities $a$, $b$, $c$ and $D_{1}$. In the pN approximation, the modified gravity sector generates further curvature dependent contributions, whereas these terms are collectively denoted by $D_{1pN}$ as well as behave as effective dark source corrections in the pN background, however their explicit form is presented as follows:}
\begin{align}
\nonumber&D_{1PN}=\frac{1}{\dot{Y} \left\{1 - \frac{n \breve{\Phi} R_o^{\,n-1}}{(1+\breve{\Upsilon} R_o^{n})^{2}}\right\}}\left[\varrho_o\left\{-Ye n\breve{\Phi} R_o^{n-2}\left[\frac{(n-1)(1+\breve{\Upsilon} R_o^{n})-2n \breve{\Upsilon} R_o^{n}}{(1+\breve{\Upsilon} R_o^{n})^{3}}\right]\right\}^{.}\right.\\\nonumber&\left.+\frac{2m_o\varrho_o\left\{-Ye n\breve{\Phi} R_o^{n-2}\left[\frac{(n-1)(1+\breve{\Upsilon} R_o^{n})-2n \breve{\Upsilon} R_o^{n}}{(1+\breve{\Upsilon} R_o^{n})^{3}}\right]\right\}^{.}}{r_{2}}+\frac{1}{2}\left\{\frac{\left\{R_o-\frac{\breve{\Phi} R_o^{n}}{1+\breve{\Upsilon} R_o^{n}}\right\}\left\{-Ye n\breve{\Phi} R_o^{n-2}\left[\frac{(n-1)(1+\breve{\Upsilon} R_o^{n})-2n \breve{\Upsilon} R_o^{n}}{(1+\breve{\Upsilon} R_o^{n})^{3}}\right]\right\}^{.}}{\left\{1 - \frac{n \breve{\Phi} R_o^{\,n-1}}{(1+\breve{\Upsilon} R_o^{n})^{2}}\right\}}\right.\right.\\\nonumber&\left.\left.-\left\{Ye\left[1-\frac{n \breve{\Phi} R_o^{\,n-1}}{(1+\breve{\Upsilon} R_o^{n})^{2}}\right]\right\}^{.}\right\}
+\frac{m_o}{r_{2}}\left\{\frac{\left\{R_o-\frac{\breve{\Phi} R_o^{n}}{1+\breve{\Upsilon} R_o^{n}}\right\}\left\{-Ye n\breve{\Phi} R_o^{n-2}\left[\frac{(n-1)(1+\breve{\Upsilon} R_o^{n})-2n \breve{\Upsilon} R_o^{n}}{(1+\breve{\Upsilon} R_o^{n})^{3}}\right]\right\}^{.}}{\left\{1 - \frac{n \breve{\Phi} R_o^{\,n-1}}{(1+\breve{\Upsilon} R_o^{n})^{2}}\right\}}\right.\right.\\\nonumber&\left.\left.-\left\{Ye\left[1-\frac{n \breve{\Phi} R_o^{\,n-1}}{(1+\breve{\Upsilon} R_o^{n})^{2}}\right]\right\}^{.}\right\}+\left\{2\dot{Y}a+2\dot{Y}b+6\frac{\dot{Y}am_o}
{r_{2}}+2\frac{\dot{Y}cm_o}{rr_{2}}+\dot{Y}c\right\}W_{00}\right.\\\nonumber&-\left\{4\frac{Ya}{r}-2\frac{Yb}{r}-\frac{Yc}{r^2}-\frac{Yc'}{r}+2\frac{\left\{-Ye n\breve{\Phi} R_o^{n-2}\left[\frac{(n-1)(1+\breve{\Upsilon} R_o^{n})-2n \breve{\Upsilon} R_o^{n}}{(1+\breve{\Upsilon} R_o^{n})^{3}}\right]\right\}}{r\left\{1 - \frac{n \breve{\Phi} R_o^{\,n-1}}{(1+\breve{\Upsilon} R_o^{n})^{2}}\right\}}\right.\\\nonumber&\left.
+4\frac{m_o\left\{-Ye n\breve{\Phi} R_o^{n-2}\left[\frac{(n-1)(1+\breve{\Upsilon} R_o^{n})-2n \breve{\Upsilon} R_o^{n}}{(1+\breve{\Upsilon} R_o^{n})^{3}}\right]\right\}}{rr_{2}\left\{1 - \frac{n \breve{\Phi} R_o^{\,n-1}}{(1+\breve{\Upsilon} R_o^{n})^{2}}\right\}}-3Ya'-2Yb'-2\frac{m_o\left\{-Ye n\breve{\Phi} R_o^{n-2}\left[\frac{(n-1)(1+\breve{\Upsilon} R_o^{n})-2n \breve{\Upsilon} R_o^{n}}{(1+\breve{\Upsilon} R_o^{n})^{3}}\right]\right\}}{r_{2}\left\{1 - \frac{n \breve{\Phi} R_o^{\,n-1}}{(1+\breve{\Upsilon} R_o^{n})^{2}}\right\}}\right.\\\nonumber&\left.+3\frac{m_o\left\{-Ye n\breve{\Phi} R_o^{n-2}\left[\frac{(n-1)(1+\breve{\Upsilon} R_o^{n})-2n \breve{\Upsilon} R_o^{n}}{(1+\breve{\Upsilon} R_o^{n})^{3}}\right]\right\}}{r^2_{2}\left\{1 - \frac{n \breve{\Phi} R_o^{\,n-1}}{(1+\breve{\Upsilon} R_o^{n})^{2}}\right\}}
-9\frac{Yam_o}{r^2_{2}}-4\frac{Ym_o}{r^2_{2}}-3\frac{Ya'm_o}{r_{2}}+6\frac{Ybm_o}{r^2_{2}}
+\frac{Ybm_o}{rr_{2}}+2\frac{Yb'm_o}{r_{2}}\right.\\\nonumber&\left.+2\frac{Ybm_o}{r^2_{2}}-4\frac{Ybm_o}{rr_{2}}\right\}W_{01}-\left\{3\frac{Ya^{\Theta}}{r^2}+3\frac{Ya^{\Theta}m_o}{r^2r_{2}}+2\frac{Yb^{\Theta}}{r^2}+\frac{Yc^{\Theta}}{r^3}-2\frac{Yb^{\Theta}m_o}{r^2r_{2}}\right\}W_{01}
+\dot{Y}bW_{11}+\frac{\dot{Y}bW_{11}}{r_{2}}+\dot{Y}bW_{22}\\\nonumber&+\frac{\dot{Y}b}{r_{2}}\times\left.W_{22}+\frac{Yc}{r}W_{33}
+2\frac{Ycm_o}{rr_{2}}W_{33}\right]-\left\{\left\{2Ya+\frac{-Ye n\breve{\Phi} R_o^{n-2}\left[\frac{(n-1)(1+\breve{\Upsilon} R_o^{n})-2n \breve{\Upsilon} R_o^{n}}{(1+\breve{\Upsilon} R_o^{n})^{3}}\right]}{\left\{1 - \frac{n \breve{\Phi} R_o^{\,n-1}}{(1+\breve{\Upsilon} R_o^{n})^{2}}\right\}}\right.\right.\\\nonumber&\left.\left.+\frac{m_oYe n\breve{\Phi} R_o^{n-2}\left[\frac{(n-1)(1+\breve{\Upsilon} R_o^{n})-2n \breve{\Upsilon} R_o^{n}}{(1+\breve{\Upsilon} R_o^{n})^{3}}\right]}{r_{2}\left\{1 - \frac{n \breve{\Phi} R_o^{\,n-1}}{(1+\breve{\Upsilon} R_o^{n})^{2}}\right\}}\right\}\frac{W_{00}}
{\left\{1 - \frac{n \breve{\Phi} R_o^{\,n-1}}{(1+\breve{\Upsilon} R_o^{n})^{2}}\right\}}\right\}^{.}-\left\{\frac{Ye n\breve{\Phi} R_o^{n-2}\left[\frac{(n-1)(1+\breve{\Upsilon} R_o^{n})-2n \breve{\Upsilon} R_o^{n}}{(1+\breve{\Upsilon} R_o^{n})^{3}}\right]}{\left\{1 - \frac{n \breve{\Phi} R_o^{\,n-1}}{(1+\breve{\Upsilon} R_o^{n})^{2}}\right\}}\right.\\\nonumber&\left.+2Ya+\frac{m_o\left\{Ye n\breve{\Phi} R_o^{n-2}\left[\frac{(n-1)(1+\breve{\Upsilon} R_o^{n})-2n \breve{\Upsilon} R_o^{n}}{(1+\breve{\Upsilon} R_o^{n})^{3}}\right]\right\}}{r_{2}\left\{1 - \frac{n \breve{\Phi} R_o^{\,n-1}}{(1+\breve{\Upsilon} R_o^{n})^{2}}\right\}}\right\}\left\{\frac{W_{00}}
{\left\{1 - \frac{n \breve{\Phi} R_o^{\,n-1}}{(1+\breve{\Upsilon} R_o^{n})^{2}}\right\}}\right\}^{.}+\left\{\left\{2Ya+2\frac{m_oYa}{r_{2}}\right.\right.\\\nonumber&\left.\left.-\frac{Ye n\breve{\Phi} R_o^{n-2}\left[\frac{(n-1)(1+\breve{\Upsilon} R_o^{n})-2n \breve{\Upsilon} R_o^{n}}{(1+\breve{\Upsilon} R_o^{n})^{3}}\right]}{\left\{1 - \frac{n \breve{\Phi} R_o^{\,n-1}}{(1+\breve{\Upsilon} R_o^{n})^{2}}\right\}}+2Yb-2\frac{m_oYb}{r_{2}}\right\}\frac{W_{01}}
{\left\{1 - \frac{n \breve{\Phi} R_o^{\,n-1}}{(1+\breve{\Upsilon} R_o^{n})^{2}}\right\}}\right\}'+\left\{2Ya+2\frac{m_oYa}{r_{2}}\right.\\\nonumber&\left.-\frac{Ye n\breve{\Phi} R_o^{n-2}\left[\frac{(n-1)(1+\breve{\Upsilon} R_o^{n})-2n \breve{\Upsilon} R_o^{n}}{(1+\breve{\Upsilon} R_o^{n})^{3}}\right]}{\left\{1 - \frac{n \breve{\Phi} R_o^{\,n-1}}{(1+\breve{\Upsilon} R_o^{n})^{2}}\right\}}\right\}\left\{\frac{W_{01}}{\left\{1 - \frac{n \breve{\Phi} R_o^{\,n-1}}{(1+\breve{\Upsilon} R_o^{n})^{2}}\right\}}\right\}'
+\left\{2Ya+2\frac{m_oYa}{r_{2}}\right.\\\nonumber&\left.-\frac{Ye n\breve{\Phi} R_o^{n-2}\left[\frac{(n-1)(1+\breve{\Upsilon} R_o^{n})-2n \breve{\Upsilon} R_o^{n}}{(1+\breve{\Upsilon} R_o^{n})^{3}}\right]}{\left\{1 - \frac{n \breve{\Phi} R_o^{\,n-1}}{(1+\breve{\Upsilon} R_o^{n})^{2}}\right\}}\right\}\left\{\frac{W_{02}}{\left\{1 - \frac{n \breve{\Phi} R_o^{\,n-1}}{(1+\breve{\Upsilon} R_o^{n})^{2}}\right\}}\right\}^{\Theta}
+\left\{\left\{2Ya+2\frac{m_oYa}{r_{2}}\right.\right.\\\nonumber&\left.\left.-\frac{Ye n\breve{\Phi} R_o^{n-2}\left[\frac{(n-1)(1+\breve{\Upsilon} R_o^{n})-2n \breve{\Upsilon} R_o^{n}}{(1+\breve{\Upsilon} R_o^{n})^{3}}\right]}{\left\{1 - \frac{n \breve{\Phi} R_o^{\,n-1}}{(1+\breve{\Upsilon} R_o^{n})^{2}}\right\}}+2Yb-2\frac{m_oYb}{r_{2}}\right\}\frac{W_{02}}
{\left\{1 - \frac{n \breve{\Phi} R_o^{\,n-1}}{(1+\breve{\Upsilon} R_o^{n})^{2}}\right\}}\right\}^{\Theta}.
\end{align}
{In the similar way, within the pN regime, the adopted modified gravity model gives rise to additional curvature dependent contributions, while these corrections are collectively represented by $D_{2pN}$ as well as serve as {effective curvature source terms} in the pN background. Their complete expressions are given as follows:}
\begin{align}
\nonumber&D_{2PN}=\frac{1}{\left\{1 - \frac{n \breve{\Phi} R_o^{\,n-1}}{(1+\breve{\Upsilon} R_o^{n})^{2}}\right\}}\left[\frac{m_o\left\{-Ye n\breve{\Phi} R_o^{n-2}\left[\frac{(n-1)(1+\breve{\Upsilon} R_o^{n})-2n \breve{\Upsilon} R_o^{n}}{(1+\breve{\Upsilon} R_o^{n})^{3}}\right]\right\}}{r_{2}^2\left\{1 - \frac{n \breve{\Phi} R_o^{\,n-1}}{(1+\breve{\Upsilon} R_o^{n})^{2}}\right\}}\varrho_o\right.+\left\{\frac{-2Ye n\breve{\Phi} R_o^{n-2}\left[\frac{(n-1)(1+\breve{\Upsilon} R_o^{n})-2n \breve{\Upsilon} R_o^{n}}{(1+\breve{\Upsilon} R_o^{n})^{3}}\right]}{\left\{1 - \frac{n \breve{\Phi} R_o^{\,n-1}}{(1+\breve{\Upsilon} R_o^{n})^{2}}\right\}^{2}}\right.\\\nonumber&\times\left\{1 - \frac{n \breve{\Phi} R_o^{\,n-1}}{(1+\breve{\Upsilon} R_o^{n})^{2}}\right\}^{'}
-\frac{4m_o\left\{-Ye n\breve{\Phi} R_o^{n-2}\left[\frac{(n-1)(1+\breve{\Upsilon} R_o^{n})-2n \breve{\Upsilon} R_o^{n}}{(1+\breve{\Upsilon} R_o^{n})^{3}}\right]\right\}\left\{1 - \frac{n \breve{\Phi} R_o^{\,n-1}}{(1+\breve{\Upsilon} R_o^{n})^{2}}\right\}^{'}}{r_{2}\left\{1 - \frac{n \breve{\Phi} R_o^{\,n-1}}{(1+\breve{\Upsilon} R_o^{n})^{2}}\right\}^{2}}+\frac{8Ybm_o}{r_{2}}-\frac{2Ybm_o}{r^2_{2}}\\\nonumber&\left.\left.+\frac{2\left\{Ye n\breve{\Phi} R_o^{n-2}\left[\frac{(n-1)(1+\breve{\Upsilon} R_o^{n})-2n \breve{\Upsilon} R_o^{n}}{(1+\breve{\Upsilon} R_o^{n})^{3}}\right]\right\}}{\left\{1 - \frac{n \breve{\Phi} R_o^{\,n-1}}{(1+\breve{\Upsilon} R_o^{n})^{2}}\right\}}-\frac{Ye n\breve{\Phi} R_o^{n-2}\left[\frac{(n-1)(1+\breve{\Upsilon} R_o^{n})-2n \breve{\Upsilon} R_o^{n}}{(1+\breve{\Upsilon} R_o^{n})^{3}}\right]}{r\left\{1 - \frac{n \breve{\Phi} R_o^{\,n-1}}{(1+\breve{\Upsilon} R_o^{n})^{2}}\right\}}-\frac{4Yb}{r}\right. \right.\\\nonumber&\left.-\frac{2m_o\left\{Ye n\breve{\Phi} R_o^{n-2}\left[\frac{(n-1)(1+\breve{\Upsilon} R_o^{n})-2n \breve{\Upsilon} R_o^{n}}{(1+\breve{\Upsilon} R_o^{n})^{3}}\right]\right\}}{rr_{2}\left\{1 - \frac{n \breve{\Phi} R_o^{\,n-1}}{(1+\breve{\Upsilon} R_o^{n})^{2}}\right\}}\right\}P_{xxo}
-\frac{\left\{1 - \frac{n \breve{\Phi} R_o^{\,n-1}}{(1+\breve{\Upsilon} R_o^{n})^{2}}\right\}^{'}}{\left\{1 - \frac{n \breve{\Phi} R_o^{\,n-1}}{(1+\breve{\Upsilon} R_o^{n})^{2}}\right\}}\bar{P}_{xx}+\frac{2m_o\left\{1 - \frac{n \breve{\Phi} R_o^{\,n-1}}{(1+\breve{\Upsilon} R_o^{n})^{2}}\right\}^{'}}{r_{2}\left\{1 - \frac{n \breve{\Phi} R_o^{\,n-1}}{(1+\breve{\Upsilon} R_o^{n})^{2}}\right\}}\bar{P}_{xx}\\\nonumber&-\left\{2Yb-\frac{6Ybm_o}{r_{2}}+\frac{\left\{1 - \frac{n \breve{\Phi} R_o^{\,n-1}}{(1+\breve{\Upsilon} R_o^{n})^{2}}\right\}}{\left\{1 - \frac{n \breve{\Phi} R_o^{\,n-1}}{(1+\breve{\Upsilon} R_o^{n})^{2}}\right\}}
-\frac{2m_o\left\{1 - \frac{n \breve{\Phi} R_o^{\,n-1}}{(1+\breve{\Upsilon} R_o^{n})^{2}}\right\}}{r_{2}\left\{1 - \frac{n \breve{\Phi} R_o^{\,n-1}}{(1+\breve{\Upsilon} R_o^{n})^{2}}\right\}}\right\}P'_{xxo}+\frac{1}{r}\\\nonumber&\times\left\{\frac{2\left\{-Ye n\breve{\Phi} R_o^{n-2}\left[\frac{(n-1)(1+\breve{\Upsilon} R_o^{n})-2n \breve{\Upsilon} R_o^{n}}{(1+\breve{\Upsilon} R_o^{n})^{3}}\right]\right\}\left\{1 - \frac{n \breve{\Phi} R_o^{\,n-1}}{(1+\breve{\Upsilon} R_o^{n})^{2}}\right\}^{'}}{\left\{1 - \frac{n \breve{\Phi} R_o^{\,n-1}}{(1+\breve{\Upsilon} R_o^{n})^{2}}\right\}^{2}}
\right.
\\\nonumber&\left.-\frac{Yb\left\{1 - \frac{n \breve{\Phi} R_o^{\,n-1}}{(1+\breve{\Upsilon} R_o^{n})^{2}}\right\}^{\Theta}}{\left\{1 - \frac{n \breve{\Phi} R_o^{\,n-1}}{(1+\breve{\Upsilon} R_o^{n})^{2}}\right\}}\right\}P_{xyo}-\left\{\frac{\left\{1 - \frac{n \breve{\Phi} R_o^{\,n-1}}{(1+\breve{\Upsilon} R_o^{n})^{2}}\right\}^{\Theta}}{r\left\{1 - \frac{n \breve{\Phi} R_o^{\,n-1}}{(1+\breve{\Upsilon} R_o^{n})^{2}}\right\}}
-\frac{2m_o\left\{1 - \frac{n \breve{\Phi} R_o^{\,n-1}}{(1+\breve{\Upsilon} R_o^{n})^{2}}\right\}^{\Theta}}{rr_{2}\left\{1 - \frac{n \breve{\Phi} R_o^{\,n-1}}{(1+\breve{\Upsilon} R_o^{n})^{2}}\right\}}\right\}\bar{P}_{xy}
\\\nonumber&+\left\{\frac{2m_o\left\{1 - \frac{n \breve{\Phi} R_o^{\,n-1}}{(1+\breve{\Upsilon} R_o^{n})^{2}}\right\}}{rr_{2}\left\{1 - \frac{n \breve{\Phi} R_o^{\,n-1}}{(1+\breve{\Upsilon} R_o^{n})^{2}}\right\}}\frac{\left\{1 - \frac{n \breve{\Phi} R_o^{\,n-1}}{(1+\breve{\Upsilon} R_o^{n})^{2}}\right\}}{r\left\{1 - \frac{n \breve{\Phi} R_o^{\,n-1}}{(1+\breve{\Upsilon} R_o^{n})^{2}}\right\}}\right\}P_{zzo}+\frac{1}{2}\left\{\left\{Ye\left[1-\frac{n \breve{\Phi} R_o^{\,n-1}}{(1+\breve{\Upsilon} R_o^{n})^{2}}\right]\right\}^{'}\right.\\\nonumber&\left.+\frac{2\left\{Ye\left[1-\frac{n \breve{\Phi} R_o^{\,n-1}}{(1+\breve{\Upsilon} R_o^{n})^{2}}\right]\right\}\left\{1 - \frac{n \breve{\Phi} R_o^{\,n-1}}{(1+\breve{\Upsilon} R_o^{n})^{2}}\right\}}{\left\{1 - \frac{n \breve{\Phi} R_o^{\,n-1}}{(1+\breve{\Upsilon} R_o^{n})^{2}}\right\}}
-\frac{\left\{R_o+\frac{\breve{\Phi} R_o^{n}}{1+\breve{\Upsilon} R_o^{n}}\right\}\left\{Ye n\breve{\Phi} R_o^{n-2}\left[\frac{(n-1)(1+\breve{\Upsilon} R_o^{n})-2n \breve{\Upsilon} R_o^{n}}{(1+\breve{\Upsilon} R_o^{n})^{3}}\right]\right\}}{\left\{1 - \frac{n \breve{\Phi} R_o^{\,n-1}}{(1+\breve{\Upsilon} R_o^{n})^{2}}\right\}}-2R'_o\right.\\\nonumber&\left.-Ye n\breve{\Phi} R_o^{n-2}\left[\frac{(n-1)(1+\breve{\Upsilon} R_o^{n})-2n \breve{\Upsilon} R_o^{n}}{(1+\breve{\Upsilon} R_o^{n})^{3}}\right]+Ye'-\frac{\left\{R_o-\frac{\breve{\Phi} R_o^{n}}{1+\breve{\Upsilon} R_o^{n}}\right\}^{'}-Ye n\breve{\Phi} R_o^{n-2}\left[\frac{(n-1)(1+\breve{\Upsilon} R_o^{n})-2n \breve{\Upsilon} R_o^{n}}{(1+\breve{\Upsilon} R_o^{n})^{3}}\right]}{\left\{1 - \frac{n \breve{\Phi} R_o^{\,n-1}}{(1+\breve{\Upsilon} R_o^{n})^{2}}\right\}}\right\}\\\nonumber&-\left\{\frac{Yam_o}{r^2_{2}}+\frac{2Ybm_o}{r_{2}^2}-\frac{m_o-Ye n\breve{\Phi} R_o^{n-2}\left[\frac{(n-1)(1+\breve{\Upsilon} R_o^{n})-2n \breve{\Upsilon} R_o^{n}}{(1+\breve{\Upsilon} R_o^{n})^{3}}\right]}{r_{2}^2\left\{1 - \frac{n \breve{\Phi} R_o^{\,n-1}}{(1+\breve{\Upsilon} R_o^{n})^{2}}\right\}}\right\}W_{00}-\dot{Y}\left\{4b-\frac{4bm_o}{r_{2}}+a
+\frac{am_o}{r_{2}}+\frac{c}{r}\right\}W_{01}\\\nonumber&+\left\{3Yb'-\frac{9Yb'm_o}{r_{2}}+\frac{9Ybm_o}{r^2_{2}}+Ya
+\frac{Yam_o}{r_{2}}-\frac{Yam_o}{r^2_{2}}+\frac{Yc'}{r}-\frac{2Yc'm_o}{rr_{2}}\right.\\\nonumber&\left.-\frac{Yc}{r^2}+\frac{2Ycm_o}{r^2r_{2}}-\frac{2Yb}{r}-\frac{2Ybm_o}{r^2_{2}}+\frac{4Ybm_o}{rr_{2}}-\frac{3-Ye n\breve{\Phi} R_o^{n-2}\left[\frac{(n-1)(1+\breve{\Upsilon} R_o^{n})-2n \breve{\Upsilon} R_o^{n}}{(1+\breve{\Upsilon} R_o^{n})^{3}}\right]}{r\left\{1 - \frac{n \breve{\Phi} R_o^{\,n-1}}{(1+\breve{\Upsilon} R_o^{n})^{2}}\right\}}\right.
\\\nonumber&\left.-\frac{3m_o\left\{Ye n\breve{\Phi} R_o^{n-2}\left[\frac{(n-1)(1+\breve{\Upsilon} R_o^{n})-2n \breve{\Upsilon} R_o^{n}}{(1+\breve{\Upsilon} R_o^{n})^{3}}\right]\right\}}{r^2_{2}\left\{1 - \frac{n \breve{\Phi} R_o^{\,n-1}}{(1+\breve{\Upsilon} R_o^{n})^{2}}\right\}}
-\frac{6m_o\left\{Ye n\breve{\Phi} R_o^{n-2}\left[\frac{(n-1)(1+\breve{\Upsilon} R_o^{n})-2n \breve{\Upsilon} R_o^{n}}{(1+\breve{\Upsilon} R_o^{n})^{3}}\right]\right\}}{rr_{2}\left\{1 - \frac{n \breve{\Phi} R_o^{\,n-1}}{(1+\breve{\Upsilon} R_o^{n})^{2}}\right\}}\right\}W_{11}+\left\{4Yb^{\Theta}\right.\\\nonumber&\left.-\frac{12Yb^{\Theta}m_o}{r_{2}}+Ya^{\Theta}
-\frac{Ya^{\Theta}m_o}{r_{2}}+\frac{Yc'}{r}-\frac{4Yc'm_o}{rr_{2}}\right\}W_{12}+\left\{Yb'-\frac{3Yb'm_o}{r_{2}}-3Yb+\frac{12Ybm_o}{r_{2}}
\right.\\\nonumber&\left.-\frac{2Yb}{r}+\frac{Ye n\breve{\Phi} R_o^{n-2}\left[\frac{(n-1)(1+\breve{\Upsilon} R_o^{n})-2n \breve{\Upsilon} R_o^{n}}{(1+\breve{\Upsilon} R_o^{n})^{3}}\right]}{r\left\{1 - \frac{n \breve{\Phi} R_o^{\,n-1}}{(1+\breve{\Upsilon} R_o^{n})^{2}}\right\}}+\frac{2m_o-Ye n\breve{\Phi} R_o^{n-2}\left[\frac{(n-1)(1+\breve{\Upsilon} R_o^{n})-2n \breve{\Upsilon} R_o^{n}}{(1+\breve{\Upsilon} R_o^{n})^{3}}\right]}{rr_{2}\left\{1 - \frac{n \breve{\Phi} R_o^{\,n-1}}{(1+\breve{\Upsilon} R_o^{n})^{2}}\right\}}\right.\\\nonumber&\left.
-\frac{m_o\left\{Ye n\breve{\Phi} R_o^{n-2}\left[\frac{(n-1)(1+\breve{\Upsilon} R_o^{n})-2n \breve{\Upsilon} R_o^{n}}{(1+\breve{\Upsilon} R_o^{n})^{3}}\right]\right\}}{r^2_{2}\left\{1 - \frac{n \breve{\Phi} R_o^{\,n-1}}{(1+\breve{\Upsilon} R_o^{n})^{2}}\right\}}\right\}W_{22}
-\frac{1}{r}\left\{\frac{Yc'}{r}-\frac{2Yc'm_o}{rr_{2}}-\frac{Yc}{r}-\frac{2Ycm_o}{rr_{2}}\right.\\\nonumber&\left.\left.-2Yb+\frac{Ybm_o}{r_{2}^2}
+\frac{Ye n\breve{\Phi} R_o^{n-2}\left[\frac{(n-1)(1+\breve{\Upsilon} R_o^{n})-2n \breve{\Upsilon} R_o^{n}}{(1+\breve{\Upsilon} R_o^{n})^{3}}\right]}{r\left\{1 - \frac{n \breve{\Phi} R_o^{\,n-1}}{(1+\breve{\Upsilon} R_o^{n})^{2}}\right\}}-\frac{2m_o\left\{Ye n\breve{\Phi} R_o^{n-2}\left[\frac{(n-1)(1+\breve{\Upsilon} R_o^{n})-2n \breve{\Upsilon} R_o^{n}}{(1+\breve{\Upsilon} R_o^{n})^{3}}\right]\right\}}{rr_{2}\left\{1 - \frac{n \breve{\Phi} R_o^{\,n-1}}{(1+\breve{\Upsilon} R_o^{n})^{2}}\right\}}\right\}W_{33}\right]
\\\nonumber&
\left\{\left\{2Yb-\frac{2Ybm_o}{r_{2}}-\frac{Ye n\breve{\Phi} R_o^{n-2}\left[\frac{(n-1)(1+\breve{\Upsilon} R_o^{n})-2n \breve{\Upsilon} R_o^{n}}{(1+\breve{\Upsilon} R_o^{n})^{3}}\right]}{\left\{1 - \frac{n \breve{\Phi} R_o^{\,n-1}}{(1+\breve{\Upsilon} R_o^{n})^{2}}\right\}}-\frac{4Ybm_o}{r_{2}}+\frac{2m_o\left\{Ye n\breve{\Phi} R_o^{n-2}\left[\frac{(n-1)(1+\breve{\Upsilon} R_o^{n})-2n \breve{\Upsilon} R_o^{n}}{(1+\breve{\Upsilon} R_o^{n})^{3}}\right]\right\}}{\left\{1 - \frac{n \breve{\Phi} R_o^{\,n-1}}{(1+\breve{\Upsilon} R_o^{n})^{2}}\right\}}\right\}\right.
\\\nonumber&\left.\times\frac{W_{11}}{\left\{1 - \frac{n \breve{\Phi} R_o^{\,n-1}}{(1+\breve{\Upsilon} R_o^{n})^{2}}\right\}}\right\}'
+\left\{\left\{2Yb-\frac{2Ybm_o}{r_{2}}-\frac{Ye n\breve{\Phi} R_o^{n-2}\left[\frac{(n-1)(1+\breve{\Upsilon} R_o^{n})-2n \breve{\Upsilon} R_o^{n}}{(1+\breve{\Upsilon} R_o^{n})^{3}}\right]}{\left\{1 - \frac{n \breve{\Phi} R_o^{\,n-1}}{(1+\breve{\Upsilon} R_o^{n})^{2}}\right\}}-\frac{4Ybm_o}{r_{2}}\right.\right.\\\nonumber&\left.\left.-\frac{2m_o\left\{-Ye n\breve{\Phi} R_o^{n-2}\left[\frac{(n-1)(1+\breve{\Upsilon} R_o^{n})-2n \breve{\Upsilon} R_o^{n}}{(1+\breve{\Upsilon} R_o^{n})^{3}}\right]\right\}}{\left\{1 - \frac{n \breve{\Phi} R_o^{\,n-1}}{(1+\breve{\Upsilon} R_o^{n})^{2}}\right\}}\right\}
\frac{W_{12}}{\left\{1 - \frac{n \breve{\Phi} R_o^{\,n-1}}{(1+\breve{\Upsilon} R_o^{n})^{2}}\right\}}\right\}^{\Theta}+D_{eff}.
\end{align}

\end{document}